\documentclass[aps,prl,twocolumn,superscriptaddress]{revtex4-2}
\usepackage[colorlinks=true,citecolor=blue,linkcolor=blue,urlcolor=blue,]{hyperref}
\usepackage{graphicx, nicefrac, textcomp}
\usepackage[normalem]{ulem}
\usepackage{changes}
\usepackage{chngcntr}
\usepackage{amssymb}
\usepackage{soul}

\usepackage{color}
\usepackage{float}
\usepackage{textcomp}
\usepackage{amstext}
\usepackage{graphicx}
\usepackage{gensymb}
\usepackage{graphics}\usepackage{subfigure}\usepackage{longtable}\usepackage{pstricks}\usepackage{dcolumn}\usepackage{bm}
\usepackage{siunitx}

\begin{document}

\title{Unconventional crystal structure of the high-pressure superconductor La$_3$Ni$_2$O$_7$}

\author{P.~Puphal}
\email[]{puphal@fkf.mpg.de}
\date{September 2023}
\author{P.~Reiss}
\email[]{p.reiss@fkf.mpg.de}
\affiliation{Max-Planck-Institute for Solid State Research, Heisenbergstra{\ss}e 1, 70569 Stuttgart, Germany}
\author{N.~Enderlein}
\affiliation{Department of Physics, Friedrich-Alexander-Universit{\"a}t Erlangen-N{\"u}rnberg (FAU), 91058 Erlangen, Germany}
\author{Y.-M.~Wu}
\affiliation{Max-Planck-Institute for Solid State Research, Heisenbergstra{\ss}e 1, 70569 Stuttgart, Germany}
\author{G.~Khaliullin}
\affiliation{Max-Planck-Institute for Solid State Research, Heisenbergstra{\ss}e 1, 70569 Stuttgart, Germany}
\author{V.~Sundaramurthy}
\affiliation{Max-Planck-Institute for Solid State Research, Heisenbergstra{\ss}e 1, 70569 Stuttgart, Germany}
\author{T.~Priessnitz}
\affiliation{Max-Planck-Institute for Solid State Research, Heisenbergstra{\ss}e 1, 70569 Stuttgart, Germany}
\author{M.~Knauft}
\affiliation{Max-Planck-Institute for Solid State Research, Heisenbergstra{\ss}e 1, 70569 Stuttgart, Germany}
\author{L.~Richter}
\affiliation{Max-Planck-Institute for Chemical Physics of Solids, N\"othnitzer Straße 40, 01187 Dresden, Germany}
\author{M.~Isobe}
\affiliation{Max-Planck-Institute for Solid State Research, Heisenbergstra{\ss}e 1, 70569 Stuttgart, Germany}
\author{P.~A.~van~Aken}
\affiliation{Max-Planck-Institute for Solid State Research, Heisenbergstra{\ss}e 1, 70569 Stuttgart, Germany}
\author{H.~Takagi}
\affiliation{Max-Planck-Institute for Solid State Research, Heisenbergstra{\ss}e 1, 70569 Stuttgart, Germany}
\author{B.~Keimer}
\affiliation{Max-Planck-Institute for Solid State Research, Heisenbergstra{\ss}e 1, 70569 Stuttgart, Germany}
\author{Y.~E.~Suyolcu}
\email[]{Eren.Suyolcu@fkf.mpg.de}
\affiliation{Max-Planck-Institute for Solid State Research, Heisenbergstra{\ss}e 1, 70569 Stuttgart, Germany}
\author{B.~Wehinger}
\affiliation{European Synchrotron Radiation Facility, 71 Avenue des Martyrs, F-38043 Grenoble, France}
\author{P.~Hansmann}
\affiliation{Department of Physics, Friedrich-Alexander-Universit{\"a}t Erlangen-N{\"u}rnberg (FAU), 91058 Erlangen, Germany}
\author{M.~Hepting}
\email[]{hepting@fkf.mpg.de}
\affiliation{Max-Planck-Institute for Solid State Research, Heisenbergstra{\ss}e 1, 70569 Stuttgart, Germany}

\date{\today}

\begin{abstract}
The discovery of high-temperature superconductivity in La$_3$Ni$_2$O$_7$ at pressures above 14 GPa has spurred extensive research efforts. Yet, fundamental aspects of the superconducting phase, including the possibility of a filamentary character, are currently subjects of controversial debates. Conversely, a crystal structure with NiO$_6$ octahedral bilayers stacked along the $c$-axis direction was consistently posited in initial studies on La$_3$Ni$_2$O$_7$. Here we reassess this structure in optical floating zone-grown La$_3$Ni$_2$O$_7$ single crystals that show signs of filamentary superconductivity. Employing scanning transmission electron microscopy and single-crystal x-ray diffraction under high pressures, we observe  multiple crystallographic phases in these crystals, with the majority phase exhibiting alternating monolayers and trilayers of NiO$_6$ octahedra, signifying a profound deviation from the previously suggested bilayer structure. Using density functional theory, we disentangle the individual contributions of the monolayer and trilayer structural units to the electronic band structure of La$_3$Ni$_2$O$_7$, providing a firm basis for advanced theoretical modeling and future evaluations of the potential of the monolayer-trilayer structure for hosting superconductivity. 
\end{abstract} 

\maketitle

\textit{Introduction.—}Layered transition metal oxides manifest a plethora of quantum phenomena, including orbital ordering, complex magnetic states, and unconventional superconductivity \cite{Khomskii2014}. A prevalent structural motif in these materials is either a perovskite or a perovskite-derived crystal lattice, serving as a conducive platform for the emergence of strong electronic correlations and exotic phase behaviors \cite{Dagotto2005,Mitchell2003,Nirala2020}. An archetypal example within this class of materials is the Ruddlesden-Popper (RP) series of ruthenates \cite{Cao1999}, denoted as (Sr,Ca)$_{n+1}$Ru$_n$O$_{3n+1}$, where $n$ is the number of consecutive RuO$_6$ octahedral layers within a structural unit. The ground state of each compound in this series is determined not only by the choice of the alkaline metal ion \cite{Braden1998,Jain2017,Feldmaier2020,Trepka2022,Maeno1994,Pustogow2019}, but also by the layer index $n$. For instance, the bilayer (BL) compound Sr$_3$Ru$_2$O$_7$ realizes quantum criticality and metamagnetism \cite{Grigera2001}, whereas trilayer (TL) Sr$_4$Ru$_3$O$_{10}$ displays ferromagnetic behavior \cite{Crawford2002}.

Recently, the RP series of nickelates has experienced a resurgence of interest \cite{Hepting2023Nature}, even though investigations into the physical properties and spin and charge ordering phenomena of the $R_{n+1}$Ni$_n$O$_{3n+1}$ compounds ($R$ = rare-earth) started over three decades ago \cite{Greenblatt1997,RodriguezCarvajal1991,Torrance1992}. The catalyst for this revival of RP nickelates are recent high-pressure studies reporting superconductivity in La$_3$Ni$_2$O$_7$ \cite{Sun2023} and La$_4$Ni$_3$O$_{10}$ \cite{sakakibara2023,Li2023La4310,Zhang2023La4310,Zhu2023La4310} with a $T_c$ of around 80 and 20 K, respectively. 

Yet, the electronic configuration of the octahedrally coordinated Ni ions in the RP series is markedly distinct from that of the earlier discovered infinite-layer nickelate superconductors \cite{Li2019,Nomura2022,Hepting2021,Chen2022,Held2022}, where  Ni ions exhibit a formal 3$d^{9}$ configuration, isoelectronic and isostructural to cuprate superconductors \cite{Keimer2015,Lee2006}. Instead, Ni in La$_3$Ni$_2$O$_7$ possesses an average 3$d^{7.5}$ configuration, given the conventional RP structure with stacked BL units, assumed in previous studies \cite{Liu2022growth,Zhang1994,Greenblatt1997,Taniguchi1995,Ling1999,Voronin2001,Wu2001}. Notably, this configuration is rather close to the formal 3$d^{7}$ state in perovskite nickelates, which were proposed to exhibit a cuprate-like electronic structure when sandwiched between layers of a band gap insulator \cite{Chaloupka2008,Hansmann2009,Boris2011}. Furthermore, an average 3$d^{7.5}$ state is reminiscent of hole-doped La$_2$NiO$_4$, where 3$d^{7}$ and 3$d^{8}$ sites coexist \cite{Uchida2012,Uchida2011PRB,Uchida2011PRL}.

Nevertheless, consensus on the theoretical framework describing the high-$T_c$ superconductivity in La$_3$Ni$_2$O$_7$ has remained elusive. While most models consider the RP structure with $n =2$ and strong intra-bilayer coupling as key ingredients, the approaches for disentangling the intermixed multiorbital states in La$_3$Ni$_2$O$_7$ vary. Specifically, the roles assigned to the low-energy Ni $3d_{x^2-y^2}$, $3d_{z^2}$, and O 2$p$ orbital degrees of freedoms diverge among the models, yielding either singlet-like states with partial analogies to those of 3$d^{9}$ cuprates  \cite{Sun2023,Qin2023,Yang2023singlet,Wu2023ZRS} or different notions in other theories \cite{Luo2023,Christiansson2023,Lechermann2023,Nakata2017,Gu2023,chen2023,Shilenko2023,Lu2023Interplay,Zhang2023dimer,Lange2023}. 

Moreover, a possibly filamentary nature of the superconducting phase has been reported \cite{zhou2023filamentary}, along with a puzzling variation in the electronic transport properties among different La$_3$Ni$_2$O$_7$ samples, including a broad spectrum of superconducting onset temperatures and whether zero-resistance is reached or a residual resistance persists \cite{wang2023,zhang2023a,hou2023,Sui2023}. These inconsistencies, previously attributed to oxygen off-stoichiometries in La$_3$Ni$_2$O$_{7\pm\delta}$ and/or varying experimental conditions, underscore the need for further investigations to elucidate the fundamental properties of the material. 

In this Letter, we reassess the physical and structural properties of La$_3$Ni$_2$O$_7$ single crystals. While we observe a maximum $T_c$ of approximately 80 K for applied pressures comparable to those in Ref.~\cite{Sun2023}, our scanning transmission electron microscopy (STEM) and single crystal x-ray diffraction (XRD) experiments reveal a crystal structure composed of alternating monolayer (ML) and trilayer (TL) units of NiO$_6$ octahedra, deviating from the conventional RP structure with BL units. We contrast our calculated band structure of the ML-TL system with conventional RP nickelates and discuss possible implications of our findings for the understanding of La$_3$Ni$_2$O$_7$ and its high-$T_c$ superconductivity.

\begin{figure}[tb]
\includegraphics[width=1.0\columnwidth]{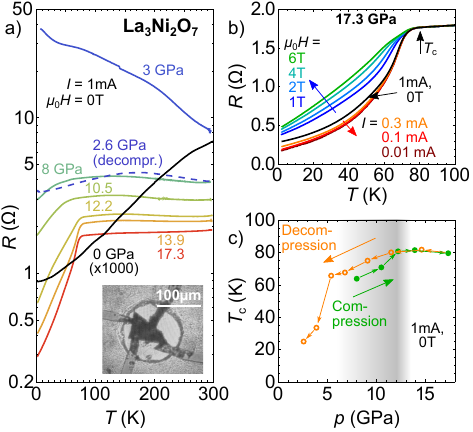}
\caption{(a) Pressure and temperature dependence of the electrical resistance of an annealed La$_3$Ni$_2$O$_7$ single crystal. Solid lines were measured upon pressure increase, while the dashed line shows the resistance after decompression to 2.6\,GPa, indicating hysteretic behavior. The inset shows the sample inside the gasket hole. (b) Field and current dependence of the resistance at the highest pressure measured. Both the field and the current are applied within the $ab$-plane. (c) Pressure-temperature phase diagram of the superconducting phase. Note the hysteresis below $p \approx 12$\,GPa.
}
\label{transport}
\end{figure}

\textit{Results.—}For the synthesis of La$_3$Ni$_2$O$_7$ single crystals we utilize the optical floating zone (OFZ) technique under high oxygen pressure. We nominally adopt the same growth conditions as reported in Refs.~\cite{Liu2022growth,Sun2023}, including an oxygen partial pressure of 15 bar (for details see Supplemental Material). Among the as-grown crystals, we observe variances in the physical properties, such as transport behaviors ranging from insulating to metallic (see Supplemental Material), which is reminiscent of previous La$_3$Ni$_2$O$_{7\pm\delta}$ powders and crystals with off-stoichiometric oxygen contents \cite{Zhang1994,Taniguchi1995,Kobayashi1996,Ling1999,Fukamachi2001,Poltavets2006,Sun2023}. A stoichiometric analysis of an as-grown crystal via inductively coupled plasma mass spectroscopy (ICP-OES) and gas extraction reveals an oxygen-deficient composition of La$_{2.98(1)}$Ni$_{1.99(1)}$O$_{6.83(7)}$. This sample will be referred to as La$_3$Ni$_2$O$_{6.83}$. On the other hand, we find that annealing of crystals at 600$^{\circ}$C in 600 bar O$_2$ atmosphere results in samples that exhibit closely consistent physical properties, including metallic transport behavior at ambient pressure [see Fig. ~\ref{transport}(a)]. Hence, we mostly focus on annealed samples, denoted as La$_3$Ni$_2$O$_7$, in the following.

The evolution of the electrical resistance of an annealed La$_3$Ni$_2$O$_7$ crystal under pressure is displayed in Fig.~\ref{transport}. We employ a Syassen-Holzapfel type diamond anvil cell (DAC) with CsI as a solid pressure medium and a conventional four-point measurement technique. Once an initial sealing pressure of approximately 3\,GPa is applied, the metallic behavior observed in  ambient-pressure measurements changes to a semiconducting-like temperature dependence [Fig.~\ref{transport}(a)]. In addition, the magnitude of the sample resistance shows a dramatic enhancement. Both observations are consistent with those reported for a subset of La$_3$Ni$_2$O$_7$ samples in previous high-pressure studies \cite{Sun2023,hou2023}. Towards higher pressure, the behavior gradually changes back to metallic, with an increasingly pronounced resistance drop at low temperatures developing from about 8\,GPa onwards. Figure~\ref{transport}(b) shows  the resistance in the region around the drop as a function of applied in-plane magnetic field and in-plane current. Evidently, with increasing field, the transition temperature is reduced (see also Supplemental Material), similarly to previous reports \cite{Sun2023}. Employing different excitation currents, we observe clear non-Ohmic behavior, but only for temperatures below the drop, demonstrating a non-metallic conductivity in this regime. Both the field and current dependence are fully consistent with the emergence of a superconducting regime below $T_c$. Moreover, a current-dependent resistance below the superconducting onset is a strong indicator for filamentary superconductivity, providing a possible explanation for the persisting residual resistance in our and several previous experiments \cite{Sun2023}.

\begin{figure*}[tb]
\includegraphics[width=2.0\columnwidth]{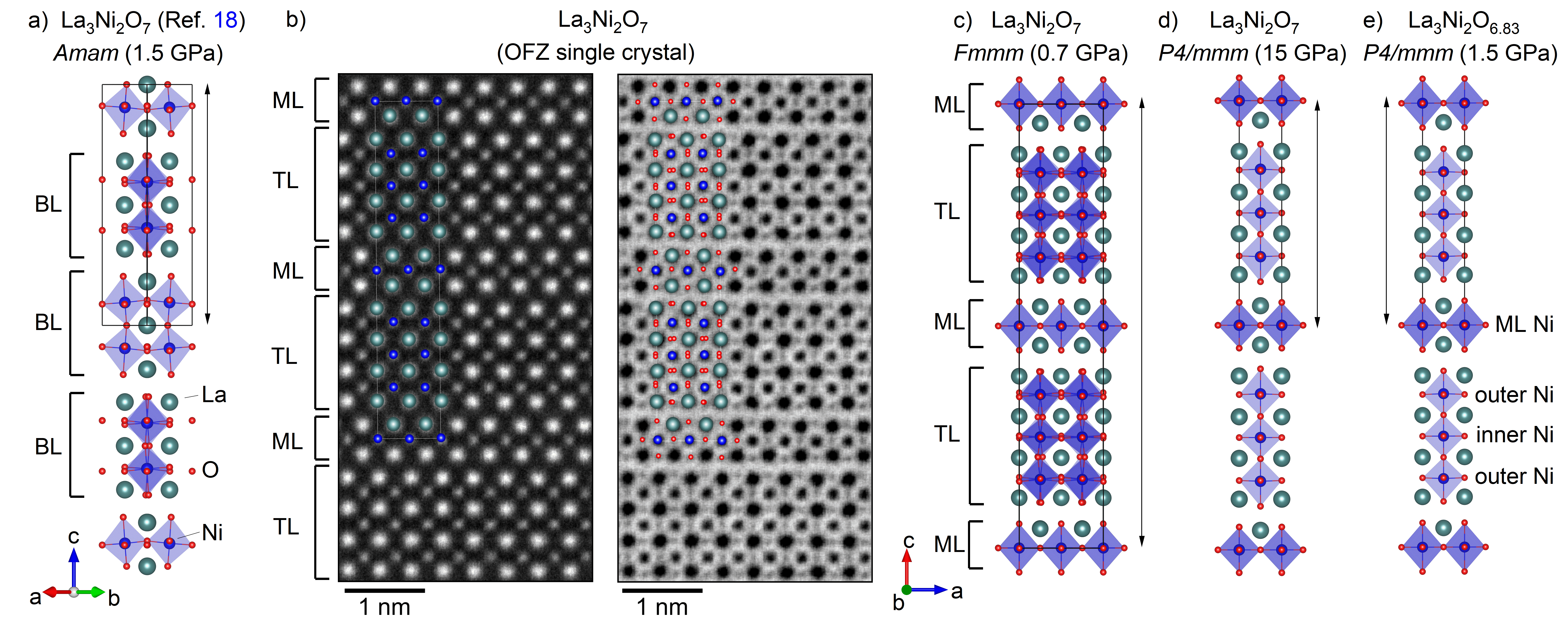}
\caption{(a) Sketch of the crystal structure of La$_3$Ni$_2$O$_7$ proposed in Ref.~\onlinecite{Sun2023}, with a sequence of BL units along the $c$ axis direction. The solid black line indicates the $Amam$ unit cell and the vertical black arrow marks the $c$ axis lattice constant. (b) Atomic-resolution STEM-HAADF (left) and STEM-ABF image (right) of our OFZ grown La$_3$Ni$_2$O$_7$ single crystal, exhibiting a stacking sequence of alternating ML and TL units. (c,d) Crystal structure of annealed La$_3$Ni$_2$O$_7$ with (c) the $Fmmm$ unit cell obtained for 0.7 GPa pressure, and (d) the $P4/mmm$ unit cell obtained for 15 GPa. Black arrows indicate the the $c$ axis lattice constants. (d). Crystal structure of oxygen-deficient La$_3$Ni$_2$O$_{6.83}$ with the $P4/mmm$ unit cell (low pressure, 1.5 GPa).
}
\label{structure}
\end{figure*}

Figure~\ref{transport}(c) summarizes the key observations of our high-pressure transport measurements. Upon compression, we identify signatures of superconductivity from $8$\,GPa onwards,  with a strong increase of $T_c$ with increasing pressure. Beyond 12\,GPa and up to 17.3\,GPa, the value of $T_c$ saturates and remains identical both upon compression and decompression. However, during decompression below 12\,GPa,  hysteretic behavior occurs, and a $T_c$  as high as 65\,K can be resolved down to 5\,GPa. Towards even lower pressures, $T_c$ suddenly drops to approximately 20\,K. Interestingly, a metallic character persists during decompression at least down to 2.6\,GPa [dashed line in Fig.~\ref{transport}(a)], which contrasts with the semiconducting behavior during compression in this pressure regime. Hence, both the superconducting and the underlying metallic phase exhibit hysteretic behavior.

Overall, the transport and magnetic properties [Fig.~\ref{transport} and Supplemental Material] of our OFZ-grown crystals are remarkably reminiscent of those reported in several previous studies
 \cite{Liu2022growth,Sun2023}, which assumed the RP structure with stacked BL units [Fig.~\ref{structure}(a)]. However, considering that other studies reported achieving a zero-resistance state below $T_c$ \cite{wang2023,zhang2023a,hou2023} and given the sensitive influence of  parameters such as oxygen content on the material's physical properties, we employ scanning transmission electron microscopy (STEM) for a more rigorous investigation of the structural nuances.

Figure~\ref{structure}(b) displays the simultaneously acquired STEM–high-angle annular dark-field (HAADF) and annular bright-field (ABF) images of an annealed La$_3$Ni$_2$O$_7$ crystal. Remarkably, although the revealed structure in Fig.~\ref{structure}(b) is layered, it consists of a sequence of alternating monolayer (ML) and trilayer (TL) blocks. Such an alternating layer stacking is a dramatic departure from the monotonous stacking of BL units [Fig.~\ref{structure}(a)], suggested in previous works on La$_3$Ni$_2$O$_7$ \cite{Liu2022growth,Zhang1994,Greenblatt1997,Taniguchi1995,Ling1999,Voronin2001,Wu2001,Sun2023,zhang2023a,hou2023,zhang2023doping,sakakibara2023,yang2023,wang2023,Hosoya2008}. Instead, the observed stacking manifests a previously unreported polytype structure \cite{Guinier1984} of the conventional nickelate RP phases. Beyond the high-resolution STEM imaging on a local scale in Fig.~\ref{structure}(b), we observe that the alternating ML-TL order persists in all investigated regions of the crystal. This suggests that the ML-TL sequence dominates the lattice of this OFZ grown crystal. 

Nevertheless, a large field-of-view STEM-HAADF survey reveals that on larger length scales the ML-TL sequence is occasionally interrupted by regions with deviating stacking orders (see complementary HAADF images in the Supplemental Material). Specifically, we observe regions with several subsequent TL units, reminiscent of the RP phase La$_4$Ni$_3$O$_{10}$, as well as a BL block interspersed between two ML units (see Supplemental Material). Yet, extended regions comprised exclusively of BL units, as would be expected according to the conventional RP phase stacking \cite{Liu2022growth,Zhang1994,Greenblatt1997,Taniguchi1995,Ling1999,Voronin2001,Wu2001,Sun2023,zhang2023a,hou2023,zhang2023doping,sakakibara2023,yang2023,wang2023,Hosoya2008}, are not detected in our STEM investigation.

\begin{figure}[tb]
\includegraphics[width=1.0\columnwidth]{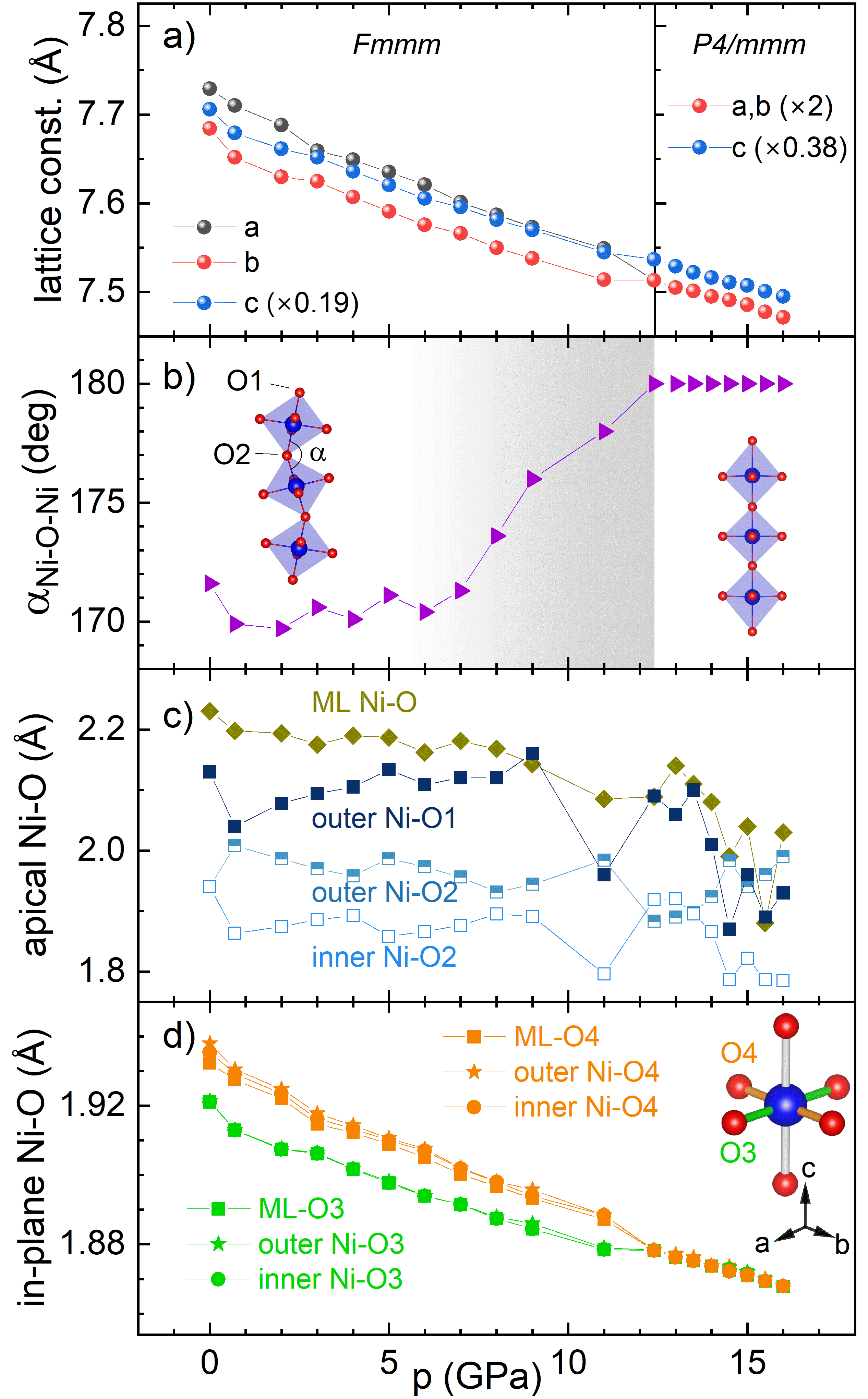}
\caption{(a) Pressure-dependence of the lattice constants of an annealed La$_3$Ni$_2$O$_{7}$, extracted from high-resolution single crystal XRD Rietveld refinement. The left and right panel demarcate the symmetry change from orthorhombic $Fmmm$ to tetragonal $P4/mmm$. (b) Ni-O-Ni bond angle along the $c$-axis direction in the TL unit. A bond angle of precisely 180$^\circ$ is intrinsic to the refinements in the $P4/mmm$ unit cell for 12.3 GPa and higher pressures. (c) Bond distances between Ni and apical O in the ML (yellow symbols) and TL unit (dark to light blue symbols). (d) Bond distances between Ni and in-plane oxygen ions, where distances along the $a$- and $b$-axis directions are color-coded in orange and green, respectively.
}
\label{pressure}
\end{figure}

To corroborate the ML-TL stacking sequence in our crystals, we turn to high-resolution synchrotron single crystal XRD. We utilize a membrane driven DAC with He as the pressure transmitting medium along with an x-ray beam focused down to a few hundred nanometers, locally probing the  pressure-dependence of the structure in  La$_3$Ni$_2$O$_7$ single crystals
Schematics of the obtained crystal structures are shown in Figs.~\ref{structure}(c)-(e) and the detailed results of the single crystal XRD refinement between 0 and 16 GPa are presented in Fig.~\ref{pressure}.

We find that the refined structures of all investigated samples (see Supplemental Material) are composed of the ML-TL sequence, consistent with our STEM observations [Fig.~\ref{structure}(b)]. For annealed La$_3$Ni$_2$O$_7$ single crystals under moderate pressure, the best refinement results are achieved when employing the orthorhombic space group $Fmmm$. The corresponding unit cell [Fig.~\ref{structure}(c)] is characterized by an essentially doubled $c$ axis lattice constant in comparison to the RP bilayer structure [Fig.~\ref{structure}(a)], which was associated with the orthorhombic space group $Amam$ in Ref.~\cite{Sun2023}. For pressures of 12.3 GPa and higher, we find that the symmetry of the ML-TL unit cell changes from $Fmmm$ to tetragonal $P4/mmm$, concomitant with a reduction of the $a,b$ and $c$ axis lattice constants [Fig.~\ref{structure}(d)]. Notably, this transition to a higher-symmetry structure coincides with the significant sharpening of the superconducting transition occurring in the transport at pressures between 12.2 and 13.9 GPa, and with the suppression of the hysteretic behavior between compression and decompression [Fig.~\ref{transport}(a) and (c)]. For comprehensiveness, we also examine the crystal structure of an as-grown  La$_3$Ni$_2$O$_{6.83}$ crystal under pressure. For this oxygen-deficient variant, the XRD refinements yield the tetragonal $P4/mmm$ unit cell already at ambient pressure [Fig.~\ref{structure}(e)], while no symmetry change occurs up to at least 20 GPa (see Supplemental Material). 

\begin{figure*}[tb]
\includegraphics[width=2.0\columnwidth]{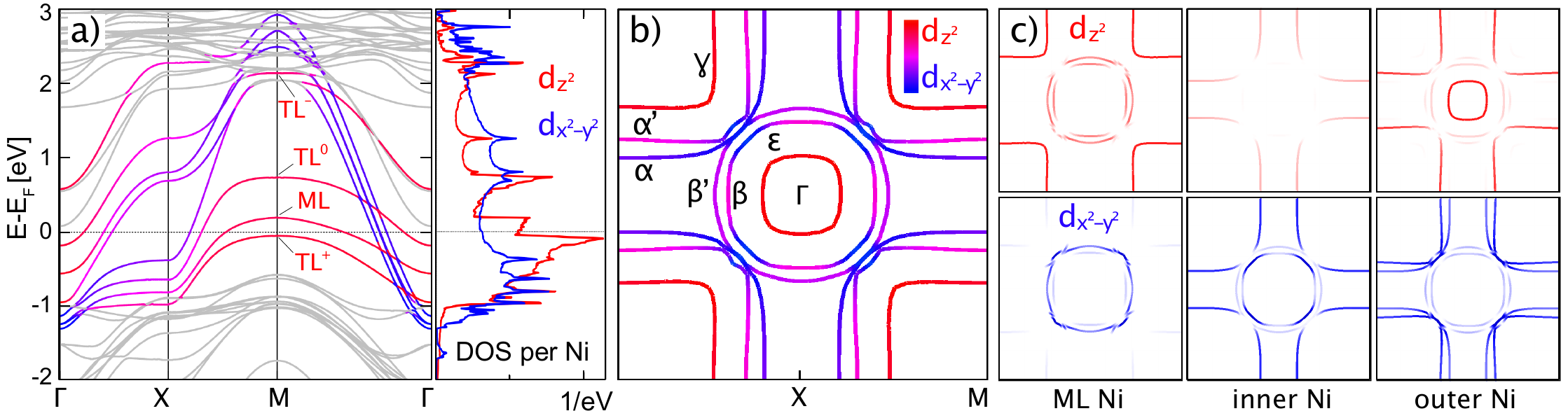}
\caption{DFT results for La$_3$Ni$_2$O$_7$ in the high-pressure $P4/mmm$ phase of the ML-TL structure. (a) The left panel shows the band structure with Ni $d_{z^2}$ and $d_{x^2-y^2}$ bands colored in red and blue, respectively, while other bands are colored in gray. The labels ML and TL indicate the structural unit from which each band originates, with the superscripts marking antibonding ($-$), nonbonding ($0$), and bonding ($+$) states around the $M$ point. The right panel shows the partial density of states (DOS). (b) FS contour plot for $k_z=0$ with four sheets ($\alpha$, $\beta$, $\gamma$, $\varepsilon$) and duplicate features ($\alpha^\prime$, $\beta^\prime$). (c) Contributions to the FS, disentangled according to the orbital states ($d_{z^2}$ or $d_{x^2-y^2}$) and Ni sites within the structural units (ML Ni, inner Ni, and outer Ni). 
}
\label{dft}
\end{figure*}

Figure~\ref{pressure}(a) displays the compression of the lattice parameters of  annealed La$_3$Ni$_2$O$_7$ with increasing pressure, progressing essentially linearly (bulk modulus $K$ = $-V dp/dV$ = 186 GPa). 
Interestingly, within the TL units, the Ni-O-Ni bond angles along the $c$ axis direction maintain a near-constant value of approximately 170$^\circ$ for pressures up to $\sim$6 GPa [Fig.~\ref{pressure}(b)], before rapidly approaching 180$^\circ$, which is reached at 12.3 GPa. This evolution of the Ni-O-Ni bond angle is qualitatively reminiscent of the increase of $T_c$ in the transport data upon compression over the same pressure range [Fig.~\ref{transport}(c)].

On the other hand, the change of the Ni-O-Ni bond angle with an onset at $\sim$6 GPa is not mirrored in the evolution of the Ni-O bond distances in the NiO$_6$ octahedra [Figs.~\ref{pressure}(c),(d)] which follow an almost linear behavior similarly to the lattice parameter in Fig.~\ref{pressure}(a). The octahedral distortions in the ML unit, particularly the asymmetry between the apical Ni-O bond length [Fig.~\ref{pressure}(c)] and the in-plane bond lengths [Fig.~\ref{pressure}(d)], are comparable to those of the ML in La$_2$NiO$_4$ \cite{Mehta1994}. For the TL unit, while the asymmetry between the outermost apical Ni-O bond and the in-plane distances is similar to that of the ML unit, its magnitude sequentially decreases for Ni-O bonds approaching the TL unit's center.

To elucidate the electronic structure of the ML-TL system, we performed calculations within density functional theory (DFT) for La$_3$Ni$_2$O$_7$ with a $P/4mmm$ unit cell and ML-TL stacking according to the XRD refinement for 12.3 GPa pressure. Figure~\ref{dft}(a) shows  the band structure and partial density of states (DOS) with a focus on the high symmetry path $\Gamma - X - M - \Gamma$ in the $k_z=0$ plane, as the dispersion along the $k_z$ direction is negligible. Consistent with previous calculations for RP nickelates with BL or TL stacking \cite{yang2023,li2017fermiology}, the states in proximity to the Fermi level are mainly of Ni $3d$ character, which we resolve into 3$d_{x^2-y^2}$ (blue) and 3$d_{z^2}$ (red) contributions, as well as the gradual mixing between these orbital characters. An exception around the $\Gamma$ point is the first band above the Fermi level, which exhibits strong La $5d$ character. Furthermore, O $2p$ character is admixed to most Ni bands, although the major spectral weight of the O $2p$ DOS is situated more than 2 eV below the Fermi level (see Supplemental Material).

From $\Gamma$ to $M$ along the diagonal $k_x=k_y$ direction, the orbital characters segregate into almost pure $d_{x^2-y^2}$ and $d_{z^2}$ states. In addition, these Ni bands can be disentangled according to their origins in specific structural units. For the case of $d_{z^2}$ bands, we highlight an isolated ML band at the $M$ point in Fig.~\ref{dft}(a), along with three bands originating from the TL unit, corresponding to bonding $(+)$, non-bonding $(0)$, and anti-bonding $(-)$ linear combinations. Notably, the TL$^+$ band resides just below the Fermi level at the $M$ point, reminiscent of the $d_{z^2}$ flat bands identified in prior DFT studies of RP nickelates with either BL or TL stacking \cite{yang2023,li2017fermiology}. The exact positioning of this band, however, is sensitive to the inclusion of a mean-field $U$ parameter in Refs.~\cite{yang2023,li2017fermiology}.

Figure~\ref{dft}(b) displays the Fermi surface (FS) in the $k_z=0$ plane, color-coded according to the contributions of $d_{x^2-y^2}$ and $d_{z^2}$ states. Four distinct features in the FS are: two circular pockets ($\beta$ and $\varepsilon$) enclosing $\Gamma$, along with square-shaped ($\gamma$) and rounded ($\alpha$) pockets centered around $M$.  Moreover, we observe the duplication of the features $\alpha$ and $\beta$, labelled as $\alpha^\prime$ and $\beta^\prime$, respectively. The $\alpha$, $\beta$, and $\gamma$ sheets in Fig.~\ref{dft}(b) are qualitatively similar to those observed for La$_4$Ni$_3$O$_{10}$ with TL stacking \cite{li2017fermiology}, while a duplication of FS features ($\alpha^\prime$ and $\beta^\prime$) was also reported in calculations for La$_3$Ni$_2$O$_{7}$ with BL stacking \cite{yang2023}. In the latter calculation, the $\gamma$ sheet is pushed below the Fermi level as a result of the inclusion of a mean-field $U$. In consequence, the $\varepsilon$ sheet around $\Gamma$ appears to be the distinguishing feature of the FS of the ML-TL system.  

This unique sheet exhibits an almost pure $d_{z^2}$ orbital character [Fig.~\ref{dft}(b)] and originates from the TL$^0$ band of the outer TL Ni layer [Fig.~\ref{dft}(c)]. In contrast, the $\beta$, $\beta^\prime$ and $\alpha$, $\alpha^\prime$ sheets generally exhibit a mixed orbital character and contributions from both structural units, except for when moving along the diagonal $k_x=k_y$ direction in Fig.~\ref{dft}(b). Besides the $\varepsilon$ sheet, also the $\gamma$ sheet with $d_{z^2}$ character is quite isolated, originating solely from the ML unit. 
Notably, the weights of the $d_{x^2-y^2}$ and $d_{z^2}$ orbital states arising from the ML unit in Fig.~\ref{dft}(c) are distributed differently from the corresponding FS features of the ML in hole-doped $R_2$NiO$_4$, which exhibits a small $d_{z^2}$-derived electron pocket around $\Gamma$ and a large hole FS of $d_{x^2-y^2}$ character, along with a high-energy pseudogap of the same symmetry as that of cuprates \cite{Uchida2011PRB}.

\textit{Discussion.—}The findings of our study carry several profound implications. Firstly, the manifestation of the ML-TL stacking in La$_3$Ni$_2$O$_7$ single crystals is remarkable in its own right. This pattern with alternating units from different RP phases extending over macroscopic length scales is distinct from common intergrowth phenomena in RP crystals, where micrometer-sized lamellae of each phase grow atop of each other \cite{Fittipaldi2007}. A rare previous example of unconventional stacking akin to the ML-TL order of La$_3$Ni$_2$O$_7$ was observed in iridates, where for carefully chosen growth conditions neither the Sr$_2$IrO$_4$ nor Sr$_3$Ir$_2$O$_7$ phase forms, but a single crystal with alternating ML-BL stacking \cite{Kim2022}.

More commonly, systems with structurally distinct building blocks, including RP phases, have previously been realized in thin film samples grown by layer-by-layer deposition techniques \cite{Wrobel2018,Mundy2016}. Yet, thin films with RP phases often suffer from stacking faults, and achieving atomically sharp interfaces in ``superlattices'' composed of distinct building blocks remains a significant challenge \cite{Suyolcu2020}. This is particularly evident in LaNiO$_3$-LaAlO$_3$ superlattices, where pronounced chemical intermixing across interfaces occurs in samples nearing the ML limit \cite{Wrobel2017}, which were initially conceptualized as a model system for modified $3d^7$ nickelates that exhibit a single-sheet $d_{x^2-y^2}$ FS \cite{Chaloupka2008,Hansmann2009}, mimicking that of superconducting cuprates. 

In our La$_3$Ni$_2$O$_7$ crystals, the interfaces between the ML and TL structural units are atomically sharp and without visible lattice defects [Fig.~\ref{structure}(b)], but a multiorbital character pervades most of the FS components [Figs.~\ref{dft}(b),(c)]. Hence, the ML-TL architecture is not akin to long-sought nickelate superlattice systems aiming to realize a single-sheet $d_{x^2-y^2}$ FS. Nonetheless, the ML unit is structurally [Figs.~\ref{structure}(b)-(e)] and electronically [Fig.~\ref{dft}] quite isolated, offering a compelling point of comparison with the ML in $R_2$NiO$_4$ \cite{Uchida2012,Uchida2011PRB,Uchida2011PRL}.
Accordingly, in spite of certain similarities to nickelates with the conventional RP stacking, the electronic structure of the ML-TL system stands out as unique. To accurately classify it within the diverse array of nickelates, cuprates, and related material systems, theoretical investigations beyond DFT will be required, complemented by experimental data, for instance from photoemission spectroscopy.

In summary, we have revealed that the ML-TL structure manifests as the primary bulk phase in our OFZ-grown La$_3$Ni$_2$O$_7$ single crystals, whereas the conventional RP phase with BL stacking was not detected. However, we also observed minority phases with stacking architectures other than ML-TL and BL, suggesting that distinct admixtures of various stacking orders might be present in individual crystals, which could be a key factor causing the previously reported sample-to-sample variations, in addition to the known oxygen off-stoichiometries. 

Utilizing a range of experimental techniques, we identified notable correlations between electrical transport properties and structural parameters, especially between the pressure dependence of $T_c$ [Fig.~\ref{transport}(c)] and the evolution of the Ni-O-Ni bond angle [Fig.~\ref{structure}(b)]. In conjunction with our observation of a filamentary character of the superconducting phase, this might imply that the superconducting filaments are either in close contact or embedded within the ML-TL bulk structure. Additionally, the dramatic increase of the normal state resistance at high pressures compared to the ambient-pressure behavior could suggest that also the observed metallic conductivity originates from filamentary channels only. Yet, in such scenarios, the origin of any filamentary nature remains enigmatic within the scope of our study. Also, it cannot be fully ruled out that an undetected BL minority phase in our crystals is responsible for the observed superconducting properties.  Nevertheless, beyond the ML-TL bulk phase and the previously proposed BL structure, our results offer additional intriguing possibilities for the superconducting host phase, which call for future explorations using advanced theoretical models and  spectroscopic methods. In particular, the superconductivity in La$_3$Ni$_2$O$_7$ crystals and powders may reside in a minority phase composed of a specific combination of ML, BL, and TL units, or arise at interfaces between these units.

\begin{acknowledgements}
We thank S. Hammoud for carrying out the ICP-OES measurement. 
The authors gratefully acknowledge the scientific support and HPC resources provided by the Erlangen National High Performance Computing Center (NHR@FAU) of the Friedrich-Alexander-Universität Erlangen-Nürnberg (FAU). The hardware is funded by the German Research Foundation (DFG). Author’s note: During the submission of this manuscript, we became aware of a report that observed a similar ML-TL stacking order in OFZ-grown La$_3$Ni$_2$O$_7$ crystals (arXiv:2312.06081).
\end{acknowledgements}

\bibliography{Literature}

\newpage

\newpage

\begin{widetext}

\newpage

\subsection*{\large Supplementary Information for ``Unconventional crystal structure of the high-pressure superconductor La$_3$Ni$_2$O$_7$''}

\subsection*{Supplementary Note 1: Optical floating zone growth at 15 bar oxygen pressure}

For the optical floating zone (OFZ) growth of La$_3$Ni$_2$O$_{7}$, we dried La$_2$O$_3$ powder (99.99\% Alfa Aesar) in a box furnace at 1100$^\circ$C. Subsequently, the precursor powder was prepared by mixing La$_2$O$_3$ and NiO (99.998\% Alfa Aesar) according to a 3:2 stoichiometry of La:Ni. The mixture was ball-milled for 20 min and transferred in alumina crucibles to a box furnace, followed by heating to 1100$^{\circ}$C for 24 hours. 
Cylindrically shaped feed and seed rods were prepared by ball-milling of the sintered materials, which were filled into rubber forms with 6~mm diameter. The rubber was evacuated and pressed in a stainless steel form filled with water using a Riken type S1-120 70 kN press. All rods were heat treated at 1150$^{\circ}$C. 

The single-crystal growth was carried out in a high pressure, high-temperature OFZ furnace (model HKZ, SciDre GmbH, Dresden, Germany), that allows for gas pressures in the growth chamber up to 300 bar. The growth chamber (sapphire single crystal) has a length of 72~mm and a wall thickness of 20~mm. A Xe arc lamp operating at 5 kW was used as a heating source within the vertical mirror alignment of the HKZ. The 14 cm feed and 4 cm seed rods were then aligned in the HKZ on steel holders followed by the installation of the high pressure chamber. Subsequently, the chamber was pressurized with 15 bar oxygen gas and held at a flow rate of 0.1 l/min. After connecting the molten zone, the growth was carried out by moving the seed with a speed of 2 mm/h. We found that this growth carried out at 15 bar oxygen partial pressure yields single crystals with a stoichiometry of La$_3$Ni$_2$O$_{7-x}$ and alternating monolayer (ML) trilayer (TL) stacking [see Figs. 2 and 3 in the main text]. 

\subsection*{Supplementary Note 2: Powder XRD of pulverized single crystals}

Supplementary Fig.~\ref{fig:XRD} displays powder x-ray diffraction (XRD) data acquired from an as-grown crystal from our growth at 15 bar oxygen pressure (see Supplementary Note 1), denoted as La$_3$Ni$_2$O$_{7-x}$ due an undetermined level of oxygen deficiency. The powder XRD analysis of pulverized single crystals provides
complementary information to the single crystal XRD refinement presented in Fig.~3 of the main text, especially in elucidating subtle differences between the ML-TL and bilayer (BL) Ruddlesden-Popper (RP) stacking in  La$_3$Ni$_2$O$_{7}$. In accord with our single crystal XRD refinement of a La$_3$Ni$_2$O$_{6.83}$ crystal (see Supplementary Note 8), the powder XRD data in Supplementary Fig.~\ref{fig:XRD}(a) can be well refined assuming the ML-TL structure of La$_3$Ni$_2$O$_{7}$ in space group $P4/mmm$. The results of the refinement, including the obtained lattice constants and atomic coordinates, are presented in Supplementary Table~\ref{tableII}(a). Notably, a refinement assuming the BL structure of La$_3$Ni$_2$O$_{7}$ in space group $Amam$ yields an almost equivalently commendable fit [Supplementary Fig.~\ref{fig:XRD}(b) and Supplementary Table~\ref{tableII}(b)] even though at least three Bragg reflexes remain unaccounted for [see black arrows in Supplementary Fig.~\ref{fig:XRD}(a)]. Hence, the differentiation between the presence of the ML-TL and BL structures in an La$_3$Ni$_2$O$_{7}$ sample can be challenging when relying solely on powder XRD data. In particular, a small admixture of either of the structures within a La$_3$Ni$_2$O$_{7}$ sample may remain undetected, owing to the small relative intensity of the few characteristic peaks of the ML-TL structure, mostly occurring at low $2\Theta$ angles. 
For comprehensiveness, we also test a refinement of our powder XRD data in the TL structure [Supplementary Fig.~\ref{fig:XRD}(c)], in correspondence to La$_4$Ni$_3$O$_{10}$ with space group $Bmab$. However, such a refinement clearly fails to describe our powder XRD data adequately.

\subsection*{Supplementary Note 3: Electrical transport and magnetization at ambient pressure }

Supplementary Fig.~\ref{fig:transport}(a, upper panel) shows two representative electrical transport curves from as-grown crystals, extracted from the boule of the 15 bar growth (see Supplementary Note 1). The as-grown La$_3$Ni$_2$O$_{7-x}$ crystals exhibit an undetermined level of oxygen deficiency. The insulating behavior observed for crystal 1 manifests in a slight resistance upturn at low temperatures, whereas crystal 2 displays metallic behavior, albeit with distinct kink-like features in the resistance curve around 160 K and 110 K. A subtle change in the slope of the metallic resistance curve occurs around 40 K. Prior to normalization,  the value of the absolute resistance at high temperatures of the insulating crystal was only marginally larger than that of the metallic crystal. Given the inherent oxygen transport gradient towards the boule center during OFZ growth \cite{Puphal2023APL,Wang2018,Zheng2020}, the contrasting resistivity behaviors of crystal 1 and 2 can be attributed to varying oxygen deficiency levels in the crystals stemming from different regions of the boule. 

Supplementary Fig.~\ref{fig:transport}(a, bottom panel) displays the magnetic susceptibility of an as-grown La$_3$Ni$_2$O$_{7-x}$ crystal, measured along different crystallographic directions and for field cooling (open symbols) and zero-field cooling (filled symbols), respectively. 
The curves measured with the field applied along an in-plane direction exhibit several anomalies, with the first one around 150 K occurring at a similar temperature as kink feature (160 K) in the resistivity. The magnetic susceptibility for fields applied along the c-axis shows the onset of an upturn at this temperature. Furthermore, at 100 K, a bifurcation of the field cooled (FC) and zero-field cooled (ZFC) curves is apparent for applied in-plane fields. An additional kink around 40 K is also present, coinciding with the subtle change in the slope of the resistivity. Note that the same crystal used for the magnetic susceptibility measurements was subsequently crushed and investigated via powder XRD [Supplementary Fig.~\ref{fig:XRD}], revealing a phase-pure composition.

Next, we anneal La$_3$Ni$_2$O$_{7-x}$ crystals under high oxygen pressure. This annealing notably results in consistent electronic transport characteristics (all annealed crystals are metallic) and mitigates the occurrence of kink-like features in the transport 
[Supplementary Fig.~\ref{fig:transport}(b), top panel]. In the magnetic susceptibility, we find that after annealing the 100 K transition is not visible any more, but the bifurcation of FC and ZFC curves around 40 K remains. One recurring feature in all crystals is the maximum in susceptibility centered around 300 K, reminiscent of that of low dimensional quantum magnets with a Heisenberg square lattice.

Supplementary Fig.~\ref{fig:comparison} compares electrical transport and magnetization of our single crystals (as-grown and annealed) to existing literature on OFZ grown La$_3$Ni$_2$O$_{7}$ crystals. Strikingly, the temperature evolution of the susceptibility [Supplementary Fig.~\ref{fig:comparison}(a)], including an anomaly at approximately 150 K and a slope change at 20 K occur similarly in our ML-TL crystal and the crystal in Ref.~\cite{Liu2022growth}, where the BL stacking was suggested. In the electrical transport [Supplementary Fig.~\ref{fig:comparison}(b)], we find that the two kinks of the metallic crystal occur at temperatures comparable to those reported for supposedly BL La$_3$Ni$_2$O$_{7}$ crystals in Ref.~\cite{Liu2022growth}, with a kink at 153 K assigned to a spin density wave (SDW) transition and a kink at 110 K assigned to a charge density wave (CDW). In the case of our annealed crystal [Supplementary Fig.~\ref{fig:comparison}(c)], the metallic resistance curve is closely reminiscent of the resistance reported for a BL La$_3$Ni$_2$O$_{7}$ crystal at ambient pressure in Ref.~\cite{Sun2023}.

\subsection*{Supplementary Note 4: High-pressure electrical transport}

For the generation of (quasi-)hydrostatic pressures for the electrical transport measurements, we employed a Syassen-Holzapfel type opposing diamond anvil cell together with a composite gasket. Using culet diameters of 400 $\micro$m, a BeCu gasket was pre-indented to a thickness of approx. 30 $\micro$m. Next, the innermost 500 $\micro$m were drilled out and shallow grooves were milled into the girdle area of the gasket. After heat treatment, a mixture of Al$_2$O$_3$ powder and Stycast 1266 was applied to the grooves to act as insulation. Additionally, a saturated, fully pre-cured mixture was pressed into the culet area to achieve the composite gasket with a final thickness of approx. 25 $\micro$m. A central hole of 175 $\micro$m diameter was drilled to produce the pressure chamber. We employed CsI as pressure transmitting medium, and we used the ruby fluorescence technique at room temperatures to determine the pressure inside the cell. Pressures were monitored before and after the low-temperature measurements and were usually found to agree within 0.2 GPa. Electrical contacts were obtained by lying four 10 $\micro$m thin Au wires over the sample. The cell was closed by adding a thin layer of CsI to the upper anvil before applying any load in order to ensure that the sample and Ruby were fully surrounded by pressure medium.

\subsection*{Supplementary Note 5: Superconducting transition}

To extract the onset of the superconducting transition at a temperature $T_c$, we employ a procedure as detailed in Supplementary Fig.~\ref{fig:TransportSM}. For any given pressure, the onset can be estimated (at least in principle) roughly from the raw data already, as shown in Supplementary Fig.~\ref{fig:TransportSM}(a-c). For a quantitative comparison, we define three different measures: (1) the onset of curvature of the zero-field, high-current data upon cooling will be denoted $T_c$, (2) the onset of deviations between the high-field data and the zero-field data will denoted $T_c^{\text{(B)}}$, and (3) the onset of deviations between the high-current data and the low-current data will be denoted $T_c^{\text{(I)}}$. In all cases, we first compute the derivatives of the resistance versus temperature, $\partial R/\partial T$, shown in Supplementary  Fig.~\ref{fig:TransportSM}(d-f). Next, we fit tangents to the high temperature and low temperature data of the zero-field, high-current case, as indicated in panel (e). The crossing point of these tangents defines $T_c$, which we show in Fig.~1(c) of the main paper. In order to extract $T_c^{\text{(B)}}$ and $T_c^{\text{(I)}}$, we compute derivative differences in order to retain only field and current dependent signatures, as shown in Supplementary Fig.~\ref{fig:TransportSM}(g-i). Again, tangents are fitted to a small temperature interval where deviations occur from zero, as indicated in Supplementary Fig.~\ref{fig:TransportSM}(h), i.e. where explicit field and current dependencies set in. This time, the crossing point of the tangent with the zero line defines the relevant $T_c^\text{(B/I)}$. 

In Supplementary Fig.~\ref{fig:TransportPhaseDiagSM}, we summarize the pressure dependence of all superconducting onset temperatures $T_c$, $T_c^{\text{(B)}}$ and $T_c^{\text{(I)}}$. Evidently, $T_c$ and $T_c^{\text{(B)}}$ trace one another almost identically whereas $T_c^{\text{(I)}}$ gives a lower onset temperature for pressures $p \gtrsim 5$ GPa. Consequently, we identify $T_c^{\text{(I)}}$ as the onset of filamentary superconductivity whereas $T_c$ indicates the formation of disconnected superconducting islands. At the lowest pressures, $T_c^{\text{(B)}} \approx T_c^{\text{(I)}}$ which most likely indicates the absence of superconducting islands. Hysteretic behavior is seen in $T_c$, $T_c^{\text{(B)}}$ and $T_c^{\text{(I)}}$ below 12 GPa. For $p = 8$ GPa, only $T_c$ is shown since no field and current dependence was measured.

\subsection*{Supplementary Note 6: Scanning transmission electron microscopy analysis}

Annealed La$_3$Ni$_2$O$_{7}$ crystals were investigated by scanning transmission electron microscopy (STEM). Electron-transparent specimens were prepared on a Thermo Fisher Scios I FIB using the standard liftout method. The lateral dimensions of the specimens were 20 $\mu$m by 1.5 $\mu$m with thicknesses between 50 and 100 nm. The STEM images were recorded by a probe aberration-corrected JEOL JEM-ARM200F scanning transmission electron microscope equipped with a cold-field emission electron source and a probe Cs corrector (DCOR, CEOS GmbH) was used at 200 kV. STEM imaging was performed at probe semiconvergence angles of 20 mrad, resulting in probe sizes of 0.8 {\AA}. Collection angles for STEM-HAADF and ABF images were 75 to 310 and 11 to 23 mrad, respectively. To improve the signal-to-noise ratio of the  data while minimizing sample damage, a high-speed time series was recorded (2 $\mu$s per pixel) and was then aligned and summed. 

Supplementary Fig.~\ref{fig:STEM} presents a large field-of-view STEM-HAADF image, complementary to the local, high-resolution STEM-HAADF image in Fig.~1(b) of the main text. Notably, while the majority phase in Supplementary Fig.~\ref{fig:STEM} consists of the ML-TL stacking, regions with different stacking order are also observed. A region with a deviating stacking sequence is further analyzed in the STEM-EELS elemental maps in Supplementary Fig.~\ref{fig:EELS}.

\subsection*{Supplementary Note 7: High-pressure synchrotron XRD measurements}

The single crystal XRD data under pressure were collected at the beamline ID27 at the European Synchrotron Radiation Facility (ESRF). We used membrane driven diamond anvil cells (DAC) and helium as the pressure transmitting medium. The applied pressure was determined from ruby fluorescence. 

The key experimental results were obtained from an annealed La$_3$Ni$_2$O$_{7}$ crystal with a size of approximately $50 \times 30 \times 20$ $\mu$m$^3$ [see loaded DAC in Supplementary Fig.~\ref{fig:DAC}(b)], using monochromatic x-rays with a wavelength of $0.22290$ {\AA}. The collection angle was -32 to 32 to degrees, allowing for the collection of about 4000 distinct indexed Bragg reflexes at each applied pressure. The single crystal XRD refinement results of the 0.7 GPa measurement for the  ML-TL structure with the orthorhombic space group $Fmmm$ are presented in Supplementary Table~\ref{tableIII}(a), and those for 16 GPa in the tetragonal space group $P4/mmm$ are given in Supplementary Table~\ref{tableIII}(b). Representative reconstructed XRD intensity maps are displayed in Supplementary Fig.~\ref{fig:XRDmaps} and  Fig.~\ref{fig:XRDmaps2}. In the $(h, 0, l)$ map in Supplementary Fig.~\ref{fig:XRDmaps}(b), pronounced streaking between the Bragg reflexes along the $l$ direction is visible, which is compatible with the presence of occasional stacking disorder of RP-derived structural units, as also observed in the large field-of-view STEM-HAADF image [Supplementary Fig.~\ref{fig:STEM}]. In addition, two other La$_3$Ni$_2$O$_{7}$ crystals were investigated [Supplementary Fig.~\ref{fig:DAC}(b)], yielding similar XRD results (not shown here). The consistency of the results among the three crystals is remarkable, since the employed few hundred nanometre-sized x-ray beam corresponds to an extremely local probe, i.e., for spatially inhomogeneous samples and/or samples where the ML-TL structure is not the majority phase such consistency would not be expected.

Furthermore, we studied an as-grown La$_3$Ni$_2$O$_{6.83}$ crystal with a size of approximately $50 \times 30 \times 20$ $\mu$m$^3$ using monochromatic x-rays with a wavelength of $0.37851$ {\AA}. The experiment was carried out with a cryostat and the collection angle was -25 to 25, allowing for the collection of about 1000 distinct indexed Bragg reflexes at each applied pressure. The loaded DAC is shown in Supplementary Fig.~\ref{fig:DAC}(a). The refinement results for 1.5 GPa in the tetragonal space group $P4/mmm$ are presented in Supplementary Table \ref{tableI}. This oxygen-deficient crystal exhibited tetragonal $P4/mmm$ symmetry already at ambient pressure. Representative reconstructed XRD intensity maps are displayed in  Supplementary Fig.~\ref{fig:XRDmaps3} and Fig.~\ref{fig:XRDmaps4}. The application of pressure yields an almost uniform compression of the lattice parameters [Supplementary Fig.~\ref{XRDo6p83}], which persists at least up to 17 GPa. At the highest measured pressures (18.5 and 20 GPa), a plateau in the lattice parameter evolution seems to emerge, although additional studies might be required to confirm this behavior.

\subsection*{Supplementary Note 8: Complementary results from band structure calculations}
The DFT results presented in the main text were calculated on basis of the experimentally determined ML-TL structure at 12.3 GPa with $P/4mmm$ symmetry. The DFT calculations were carried out using the Quantum ESPRESSO suite \cite{giannozzi2009quantum,giannozzi2017advanced}. Optimized normconserving pseudopotentials \cite{hamann2013optimized,van2018pseudodojo} were used with an energy cutoff of 120 Ry and the exchange-correlation interaction was approximated by the Perdew-Burke-Ernzerhof functional (PBE). Brillouin zone integration for the calculation of the electronic ground state was carried out on a 10 x 10 x 2
electron momentum grid, where a Gaussian smearing of 0.02 Ry was included. The partial density of states (DOS) and Fermi surface were derived from NSCF calculations (using the converged electron density from the SCF run as input) utilizing a dense 30 x 30 x 6 electron momentum grid in combination with the optimized tetrahedron method \cite{kawamura2014improved}.

While the plot of the band structure and partial DOS in Fig.~4(a) of the main text focuses on the Ni $e_g$ states, we present a more comprehensive plot in Supplementary Fig.~\ref{fig:DFT_large-bands}. Furthermore, a fatband plot illustrating the contributions of La $d$ (orange), Ni $d$ (green), and O $p$ (red) states in a broad energy window around the Fermi level is given in Supplementary Fig.~\ref{fig:La_fatbands}. In addition, we present complementary plots of the Ni $e_g$ bands disentangled according to $d_{z^2}$ and $d_{x^2-y^2}$ orbitals and their origin from Ni sites in the different structural units in Supplementary Fig.~\ref{fig:DFT_fatbands}.

\newpage

\begin{table}[tb]
\caption{Results of the ambient-pressure powder XRD Rietveld refinement of an as-grown La$_3$Ni$_2$O$_{7-x}$ single crystal that was pulverized. (a) Atomic coordinates from the refinement in the ML-TL structure with the tetragonal space group $P4/mmm$ (\#123). The obtained lattice constants are $a,b=3.84886(14)$ {\AA} and $c=20.3391(11)$ {\AA}. The reliability factor is $\chi^2$ = 5.83. (b) Refinement in the BL structure with the orthorhombic space group $Amam$ (\#63). The obtained lattice constants are $a=5.42126(19)$ {\AA}, $b=5.45918(19)$ {\AA}, and $c=20.3416(8)$ {\AA}. The reliability factor is $\chi^2$ = 4.56.}
\begin{tabular}{llllllllllllllllll}
a) & atom  & site  & x  & y  & z  & U$_{iso}$ &  &  &  &  & b) & atom  & site  & x  & y  & z  & U$_{iso}$\tabularnewline
\hline 
 & La  & La01  & 1  & 1  & 0.4108(3)  & 0.0133(3) &  &  &  &  &  & La  & La01  & 0.25  & 0.236(6)  & 0.5  & 0.0145(3)\tabularnewline
 & La  & La02  & -0.5  & -0.5  & 0.0883(3)  & 0.0133(3) &  &  &  &  &  & La  & La2  & 0.25  & 0.257(4)  & 0.31820(10)  & 0.0145(3)\tabularnewline
 & La  & La03  & 0  & 0  & 0.2234(3)  & 0.0133(3) &  &  &  &  &  & Ni  & Ni1  & 0.25  & 0.247(7)  & 0.0963(3)  & 0.0237(12)\tabularnewline
 & Ni  & Ni04  & 0  & 0  & 0  & 0.036(2) &  &  &  &  &  & O  & O1  & 0.25  & 0.257(14)  & 0  & 0.054(5)\tabularnewline
 & Ni  & Ni05  & 0.5  & 0.5  & 0.3015(7)  & 0.036(2) &  &  &  &  &  & O  & O2  & 0.25  & 0.261(12)  & 0.1840(11)  & 0.054(5)\tabularnewline
 & Ni  & Ni06  & 0.5  & 0.5  & 0.5  & 0.036(2) &  &  &  &  &  & O  & O3  & 0  & 0.5  & 0.096(3)  & 0.054(5)\tabularnewline
 & O  & O07  & 0  & 0  & 0.109(2)  & 0.011(3) &  &  &  &  &  & O  & O4  & 0.5  & 0  & 0.094(4)  & 0.054(5)\tabularnewline
 & O  & O08  & -0.5  & 0  & 0  & 0.011(3) &  &  &  &  &  &  &  &  &  &  & \tabularnewline
 & O  & O09  & 0.5  & 0.5  & 0.196(2)  & 0.011(3) &  &  &  &  &  &  &  &  &  &  & \tabularnewline
 & O  & O10  & 0.5  & 0.5  & 0.395(3)  & 0.011(3) &  &  &  &  &  &  &  &  &  &  & \tabularnewline
 & O  & O11  & 0  & 0.5  & 0.3032(14)  & 0.011(3) &  &  &  &  &  &  &  &  &  &  & \tabularnewline
 & O  & O12  & 0.5  & 0  & 0.5  & 0.011(3) &  &  &  &  &  &  &  &  &  &  & \tabularnewline
\end{tabular}\label{tableII} 
\end{table}

\begin{table}[tb]
\caption{Results of the high-pressure single crystal XRD refinement on an annealed La$_3$Ni$_2$O$_{7}$ crystal. (a) Atomic coordinates from the refinement in the ML-TL structure with the orthorhombic space group $Fmmm$ (\#69) measured at 0.7 GPa. The obtained lattice constants are $a=7.6771(7)$ {\AA}~, $b=7.672(2)$ {\AA}, and $c=40.558(3)$ {\AA}. (b) Atomic coordinates from the refinement in the ML-TL structure with the tetragonal space group $P4/mmm$ (\#123) measured at 16 GPa. The obtained lattice constants are $a,b=3.7358(2)$ {\AA} and $19.7252(13)$ {\AA}. }

\begin{tabular}{llllllllllllll}
a) atom & site & x & y & z & U$_{iso}$ &  &  & b) atom & site & x & y & z & U$_{iso}$\tabularnewline
\hline 
La & La01 & 0.5 & 0.24966(8) & 0.29661(2) & 0.0358(5) &  &  & La & La01 & 1 & 1 & 0.40715(6) & 0.0131(3)\tabularnewline
La & La02 & 0.75053(5) & 0.5 & 0.54351(2) & 0.0348(5) &  &  & La & La02 & -0.5 & -0.5 & 0.08700(6) & 0.0133(3)\tabularnewline
La & La03 & 0.5 & 0.25149(8) & 0.38669(2) & 0.0351(5) &  &  & La & La03 & 0 & 0 & 0.22640(7) & 0.0129(4)\tabularnewline
Ni & Ni04 & 0.5 & 0.2503(2) & 0.5 & 0.0318(8) &  &  & Ni & Ni04 & 0 & 0 & 0 & 0.0132(7)\tabularnewline
Ni & Ni05 & 0.75045(11) & 0.5 & 0.34547(4) & 0.0336(7) &  &  & Ni & Ni05 & 0.5 & 0.5 & 0.30859(15) & 0.0117(5)\tabularnewline
Ni & Ni06 & 0.75 & 0.5 & 0.25 & 0.0309(7) &  &  & Ni & Ni06 & 0.5 & 0.5 & 0.5 & 0.0129(8)\tabularnewline
O & O07 & 0.5 & 0.2512(10) & 0.4462(3) & 0.0346(16) &  &  & O & O07 & 0 & 0 & 0.1029(15) & 0.016(3)\tabularnewline
O & O08 & 0.75 & 0.25 & 0.5 & 0.039(3) &  &  & O & O08 & -0.5 & 0 & 0 & 0.018(3)\tabularnewline
O & O09 & 0.5 & 0 & 0.5 & 0.032(2) &  &  & O & O09 & 0.5 & 0.5 & 0.2108(16) & 0.022(4)\tabularnewline
O & O10 & 0.5 & 0.5 & 0.5 & 0.034(2) &  &  & O & O10 & 0.5 & 0.5 & 0.4095(9) & 0.016(3)\tabularnewline
O & O11 & 0.7691(9) & 0.5 & 0.2955(2) & 0.085(8) &  &  & O & O11 & 0 & 0.5 & 0.3089(8) & 0.015(2)\tabularnewline
O & O12 & 0.7467(6) & 0.5 & 0.39649(16) & 0.0273(15) &  &  & O & O12 & 0.5 & 0 & 0.5 & 0.030(6)\tabularnewline
O & O13 & 0.75 & 0.75 & 0.3441(3) & 0.0378(19) &  &  &  &  &  &  &  & \tabularnewline
O & O14 & 0.5 & 0.5 & 0.34259(18) & 0.040(3) &  &  &  &  &  &  &  & \tabularnewline
O & O15 & 1 & 0.5 & 0.34847(18) & 0.036(2) &  &  &  &  &  &  &  & \tabularnewline
O & O16 & 0.75 & 0.75 & 0.25 & 0.105(19) &  &  &  &  &  &  &  & \tabularnewline
O & O17 & 1 & 0.5 & 0.24704(13) & 0.0415(13) &  &  &  &  &  &  &  & \tabularnewline
\end{tabular}
 \label{tableIII} 
\end{table}

\begin{table}[tb]
\caption{Results of the high-pressure single crystal XRD refinement on an as-grown La$_3$Ni$_2$O$_{6.83}$ crystal. (a) Atomic coordinates from the refinement in the ML-TL structure with the tetragonal space group $P4/mmm$ (\#69) measured at 1.5 GPa. The obtained lattice constants are $a,b=3.8332(13)$ {\AA} and $c=20.244(4)$ {\AA}. (b) Atomic coordinates from the refinement in the ML-TL structure with the tetragonal space group $P4/mmm$ (\#123) measured at 14.9 GPa. The obtained lattice constants are $a,b=3.7462(2)$ {\AA} and $c=19.7507(10)$ {\AA}. }
\begin{tabular}{llllllllllllll}
a) atom & site & x & y & z & U$_{iso}$ &  &  & b) atom & site & x & y & z & U$_{iso}$\tabularnewline
\hline 
La & La01 & 1 & 1 & 0.40588(18) & 0.0394(18) &  &  & La & La01 & 1 & 1 & 0.40697(6) & 0.0116(6)\tabularnewline
La & La02 & -0.5 & -0.5 & 0.08635(16) & 0.040(2) &  &  & La & La02 & -0.5 & -0.5 & 0.08651(7) & 0.0116(6)\tabularnewline
La & La03 & 0 & 0 & 0.22655(17) & 0.0394(19) &  &  & La & La03 & 0 & 0 & 0.22680(7) & 0.0116(6)\tabularnewline
Ni & Ni04 & 0 & 0 & 0 & 0.031(3) &  &  & Ni & Ni04 & 0 & 0 & 0 & 0.0124(10)\tabularnewline
Ni & Ni05 & 0.5 & 0.5 & 0.3088(4) & 0.033(2) &  &  & Ni & Ni05 & 0.5 & 0.5 & 0.30858(18) & 0.0105(9)\tabularnewline
Ni & Ni06 & 0.5 & 0.5 & 0.5 & 0.028(3) &  &  & Ni & Ni06 & 0.5 & 0.5 & 0.5 & 0.0103(10)\tabularnewline
O & O07 & 0 & 0 & 0.110(3) & 0.08(4) &  &  & O & O07 & 0 & 0 & 0.1086(9) & 0.032(9)\tabularnewline
O & O08 & -0.5 & 0 & 0 & 0.043(17) &  &  & O & O08 & -0.5 & 0 & 0 & 0.017(4)\tabularnewline
O & O09 & 0.5 & 0.5 & 0.206(2) & 0.047(18) &  &  & O & O09 & 0.5 & 0.5 & 0.2043(10) & 0.025(6)\tabularnewline
O & O10 & 0.5 & 0.5 & 0.4047(19) & 0.047(18) &  &  & O & O10 & 0.5 & 0.5 & 0.4054(7) & 0.020(5)\tabularnewline
O & O11 & 0 & 0.5 & 0.3123(15) & 0.043(14) &  &  & O & O11 & 0 & 0.5 & 0.3091(8) & 0.016(3)\tabularnewline
O & O12 & 0.5 & 0 & 0.5 & 0.06(4) &  &  & O & O12 & 0.5 & 0 & 0.5 & 0.018(4)\tabularnewline
\end{tabular}
\label{tableI} 
\end{table}

\begin{figure*}[h]
 \begin{centering}
\includegraphics[width=0.8\columnwidth]{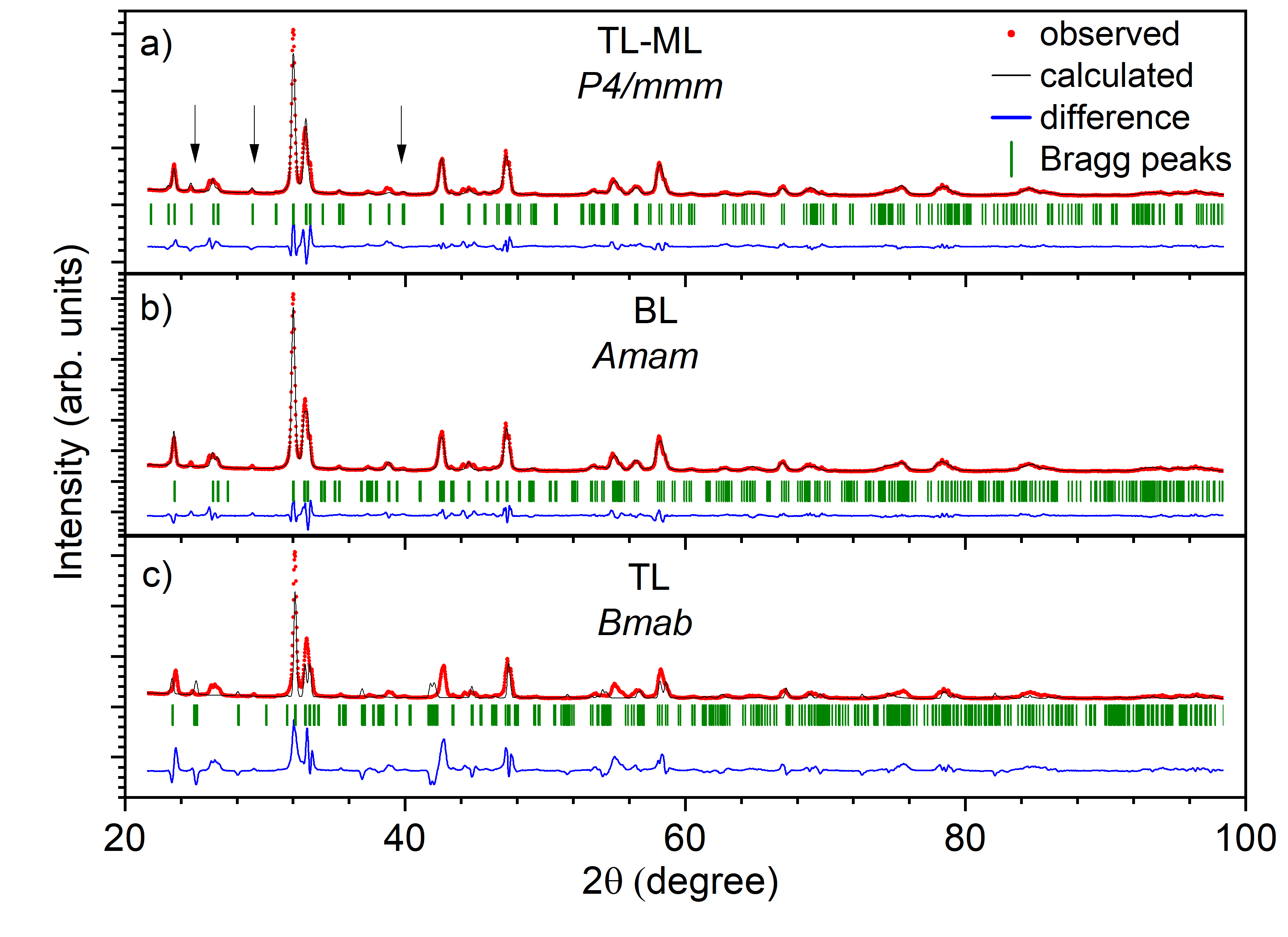}
\par\end{centering}
\caption{(a) Powder XRD pattern of a pulverized La$_3$Ni$_2$O$_{7-x}$ single crystal, acquired with Cu $K_\alpha$ radiation at ambient pressure and $T = 300$ K. The Rietveld refinement was performed for the ML-TL structure of La$_3$Ni$_2$O$_{7}$ in space group $P4/mmm$. The solid black line corresponds to the intensity calculated from the refinement, the solid blue line is the difference between the measured and calculated intensity, and the vertical green bars are the calculated Bragg peak positions. The three arrows highlight distinct peaks in the XRD data that are only captured in the ML-TL refinement. (b) The same powder XRD data, but with a refinement assuming the BL structure of La$_3$Ni$_2$O$_{7}$ in space group $Amam$. Overall, the XRD data and the calculated intensity match well, but specifically the Bragg peaks indicated by the black arrows in panel (a) are not present in the simulated BL curve in panel (b). (c) The same powder XRD data, but with a refinement assuming the TL structure of La$_4$Ni$_3$O$_{10}$ in space group $Bmab$. A strong discrepancy exists between the XRD data and the calculated intensity, excluding a significant volume fraction of pure TL stacking within the pulverized La$_3$Ni$_2$O$_{7-x}$ crystal.   
}
\label{fig:XRD}
\end{figure*}

\begin{figure*}[h]
 \begin{centering}
\includegraphics[width=0.8\columnwidth]{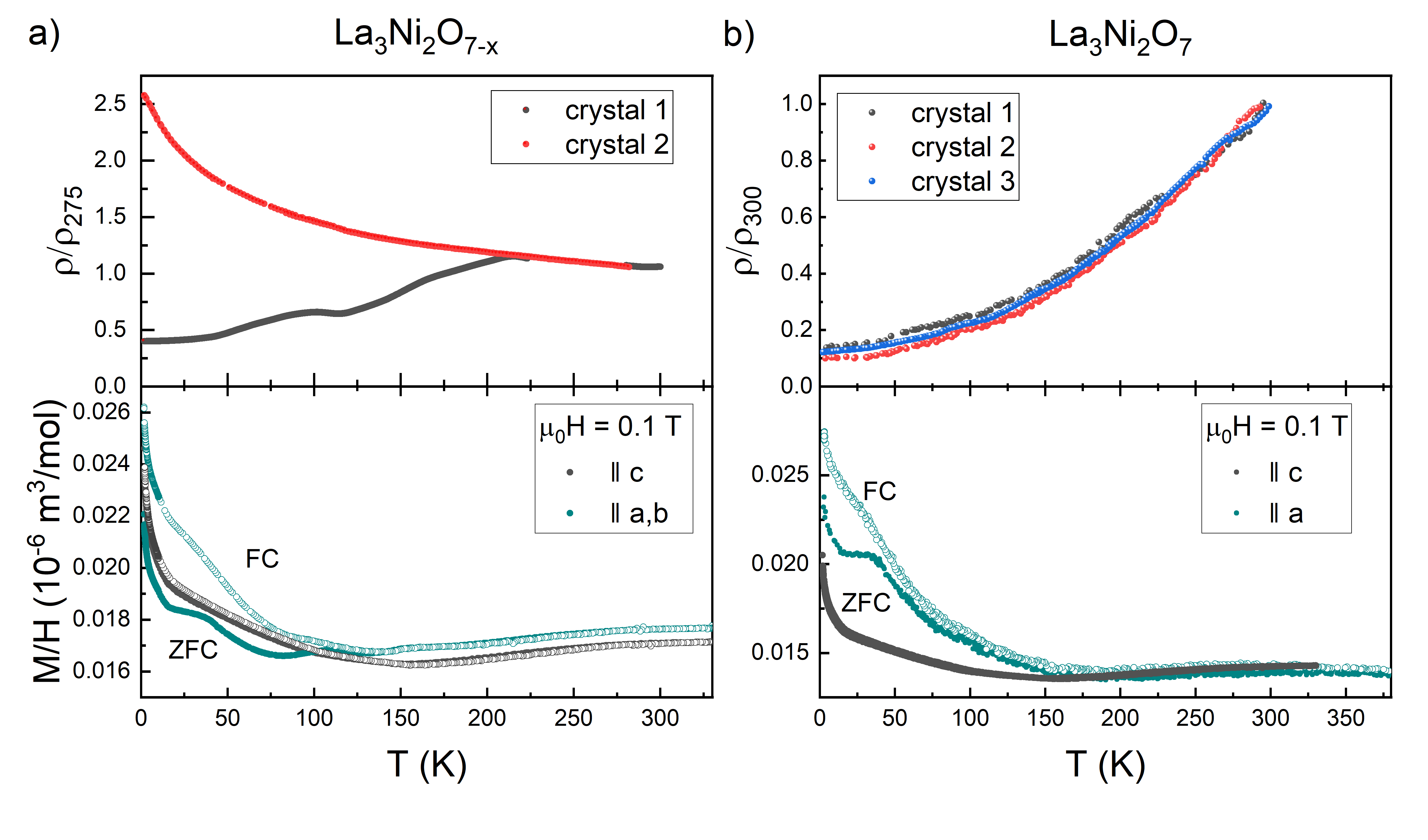}
\par\end{centering}
\caption{(a) Top panel: Electrical transport of two representative as-grown La$_3$Ni$_2$O$_{7-x}$ crystals. The resistance is normalized to 275 K for better comparability. Bottom panel: Magnetic susceptibility of an as-grown La$_3$Ni$_2$O$_{7-x}$ crystal, measured with a field of 0.1 T applied along the $c$ axis (gray) and and in-plane direction (green), respectively.  The filled symbols correspond to zero field cooled (ZFC) measurements, whereas the open symbols are field cooled (FC). (b) Top panel: Electrical transport of three different annealed La$_3$Ni$_2$O$_{7}$ crystals. Bottom panel: Magnetic susceptibility of an annealed crystal.
}
\label{fig:transport}
\end{figure*}

\begin{figure*}[h]
 \begin{centering}
\includegraphics[width=0.8\columnwidth]{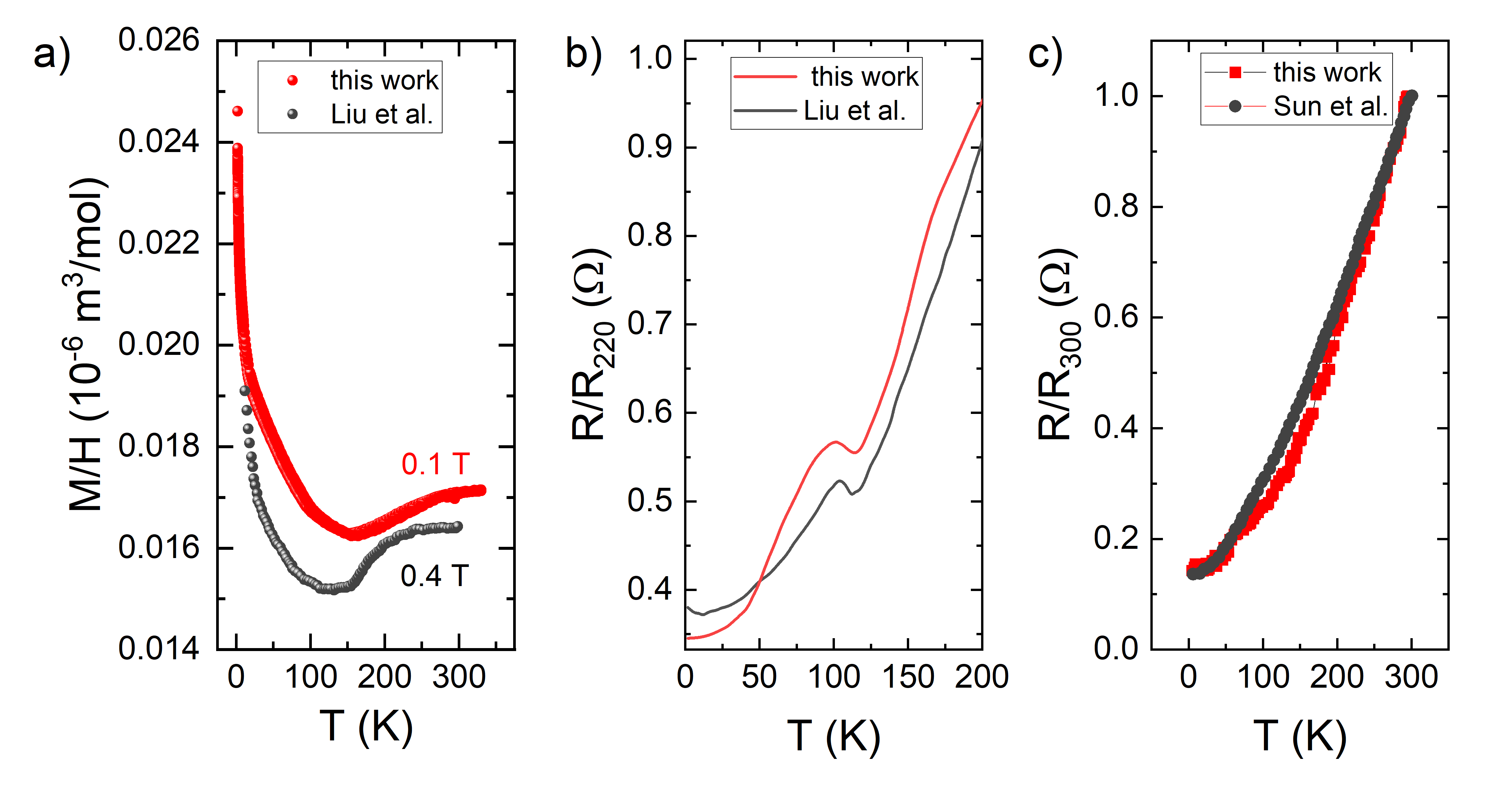}
\par\end{centering}
\caption{(a) Comparison between the $c$ axis susceptibility of our as-grown La$_3$Ni$_2$O$_{7-x}$ crystal (field 0.1 T) and an OFZ-grown crystal in Ref.~\cite{Liu2022growth} (field 0.4 T). (b) Comparison of the electrical transport of our as-grown La$_3$Ni$_2$O$_{7-x}$ crystal with an OFZ-grown crystal in Ref.~\cite{Liu2022growth}. (c) Comparison between the electrical transport of our annealed La$_3$Ni$_2$O$_{7}$ crystal and the ambient pressure data from an OFZ-grown crystal in Ref.~\cite{Sun2023}. 
}
\label{fig:comparison}
\end{figure*}

\begin{figure*}[tb]
 \begin{centering}
\includegraphics[width=0.8\columnwidth]{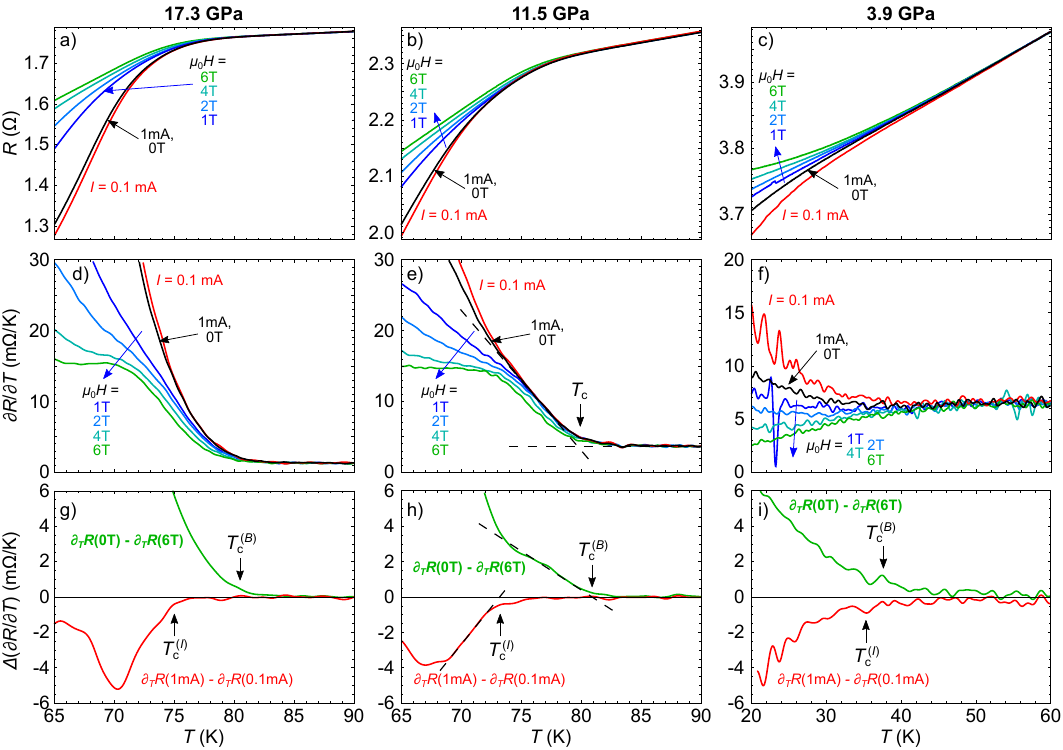}
\par\end{centering}
\caption{Extraction of the superconducting transition temperature $T_c$ from the high-pressure electrical transport measurements. (a-c) Raw resistance data for $p = 17.3$ GPa, $11.5$ GPa, and $3.9$ GPa, respectively, for various fixed in-plane currents and in-plane applied magnetic fields. (d-f) First derivative in temperature of the data shown in panels (a-c). Evidently, for high pressures, a deviation between the high-field and zero-field data occurs at a significantly higher temperature than between the high-current and low-current data. In contrast, at low pressure of 3.9 GPa, both the high-field and the low-current data deviate from the reference data around the same temperature. (g-i) Subtracting the derivatives from panels (d-f) as indicated, the onset transition temperatures can be extracted reliably. $T_c$ refers to the superconducting transition temperature as referred in the main paper, which relies only on the zero-field, high-current data. In contrast, $T_c^\text{(B)}$ refers the onset superconducting transition obtained from the analysis under different magnetic fields, $T_c^\text{(I)}$ under different currents. Dashed lines in panels (e) and (h) are linear fits to the data.}
\label{fig:TransportSM}
\end{figure*}

\begin{figure*}[tb]
 \begin{centering}
\includegraphics[width=8cm]{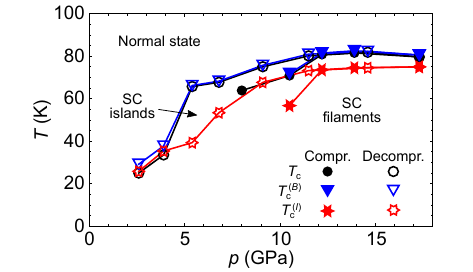}
\par\end{centering}
\caption{Temperature-pressure phase diagram of La$_3$Ni$_2$O$_7$. For the definition of the superconducting transition temperatures, see Fig.~\ref{fig:TransportSM}. At high pressure $p \gtrsim 5$ GPa, $T_c^\text{(B)} > T_c^\text{(I)}$ which suggests the formation of superconducting islands inside the intermediate temperature regime. Below $T_c^\text{(I)}$, superconducting filaments emerge. At low pressures, only weak signatures of filamentary superconductivity are observed, but no superconducting islands.}
\label{fig:TransportPhaseDiagSM}
\end{figure*}

\begin{figure*}[h]
 \begin{centering}
\includegraphics[width=0.6\columnwidth]{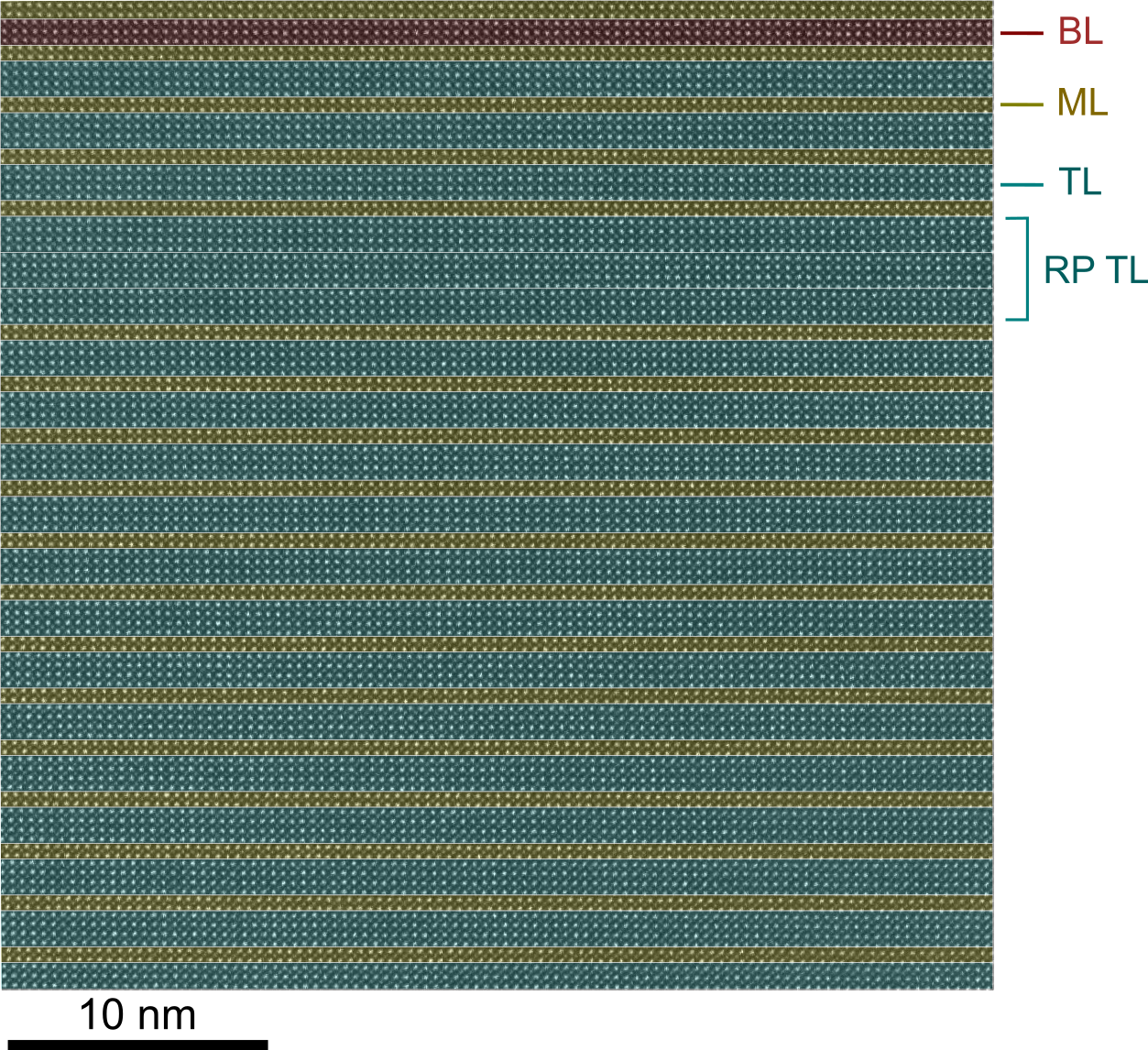}
\par\end{centering}
\caption{Large field-of-view STEM-HAADF image acquired from an annealed La$_3$Ni$_2$O$_{7}$ single crystal revealing the presence of different stacking orders of ML (yellow), BL (red), and TL (blue) units. The majority of the crystal structure is composed of a repeated ML-TL sequence, while one ML-BL-ML sequence is found at the top of the image, as well as a sequence of three TL blocks, in analogy to the RP structure with TL stacking in La$_4$Ni$_3$O$_{10}$. 
}
\label{fig:STEM}
\end{figure*}

\begin{figure*}[h]
 \begin{centering}
\includegraphics[width=.7\columnwidth]{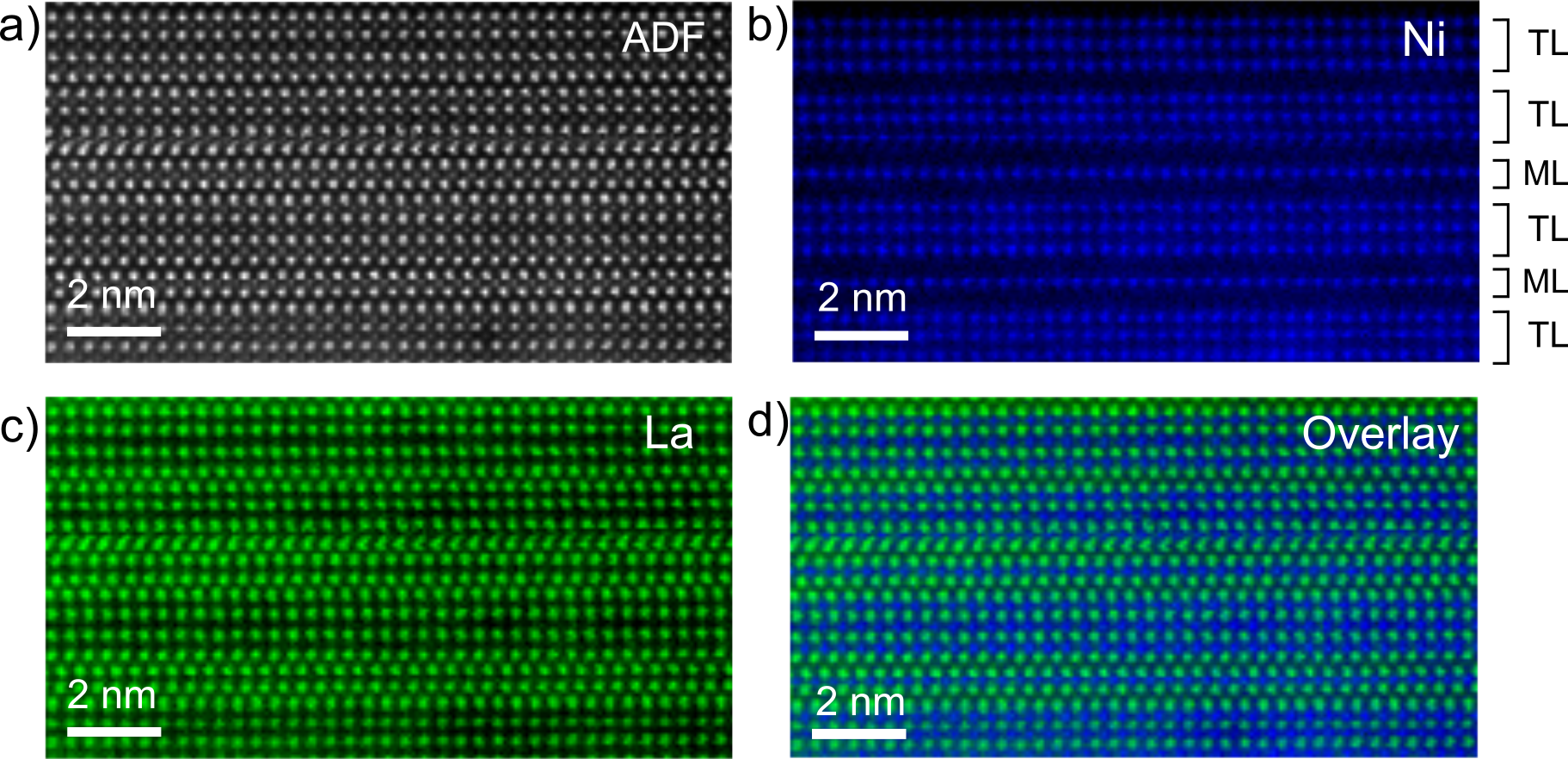}
\par\end{centering}
\caption{Atomic-resolution STEM-ADF image of an annealed La$_3$Ni$_2$O$_{7}$ single crystal (a) along with the simultaneously acquired STEM-EELS elemental maps of Ni (b, blue), La (c, green), and the overlaid map (d).  The Ni $L_{3,2}$ and La $M_{5,4}$ edges were used for the acquisition of the Ni and La maps, respectively. The stacking sequence in the investigated region is TL-TL-ML-TL-ML-TL. Hence, the topmost block deviates from a continuous ML-TL sequence. Within the atomic layers of the TL and ML blocks, the La and Ni contents are both distributed homogeneously. 
}
\label{fig:EELS}
\end{figure*}

\begin{figure*}[h]
 \begin{centering}
\includegraphics[width=0.8\columnwidth]{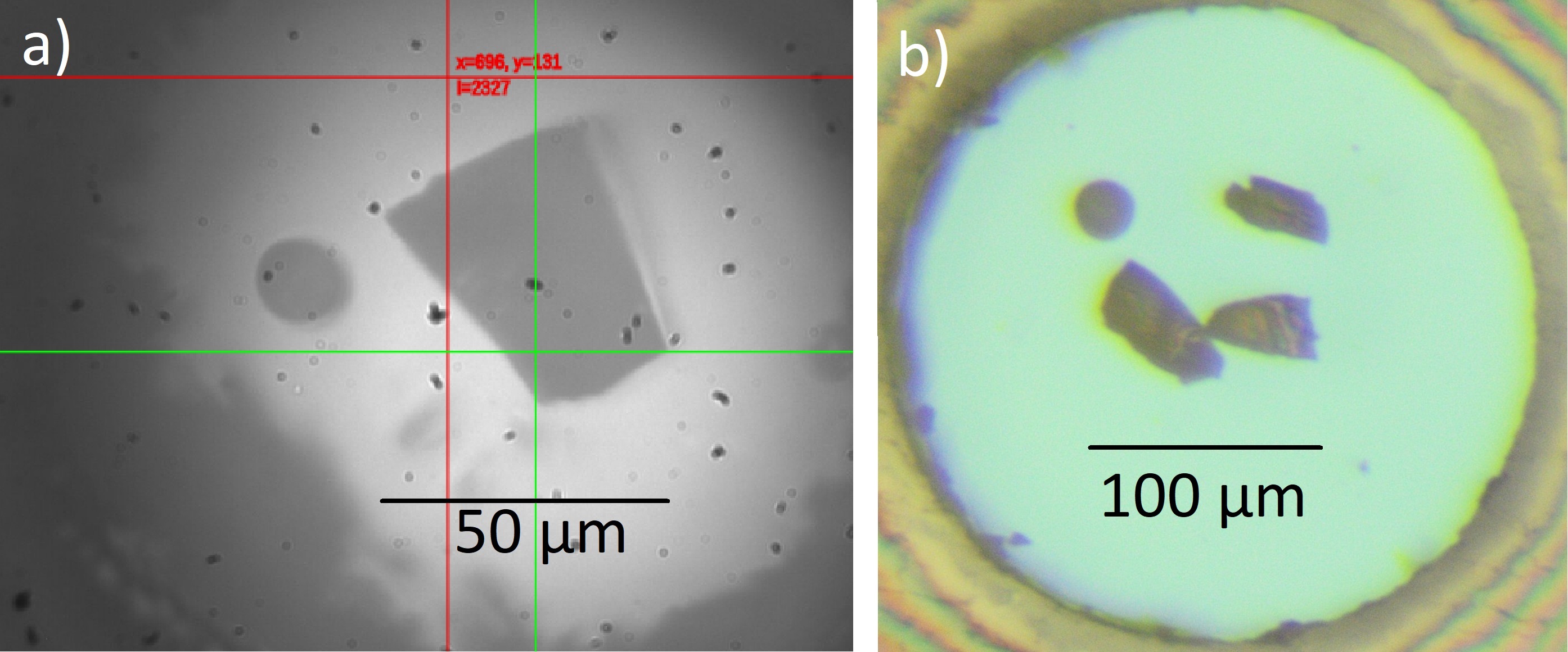}
\par\end{centering}
\caption{Pictures of the loaded DAC with a 0.5 mm culet diameter and a 0.2 mm steel gasket diameter, with (a) an as grown La$_3$Ni$_2$O$_{6.83}$ crystal and (b) three annealed La$_3$Ni$_2$O$_{7}$ crystals. The spherical object is a ruby crystal used for reference measurements for pressure calibration.
}
\label{fig:DAC}
\end{figure*}

\begin{figure*}[h]
 \begin{centering}
\includegraphics[width=1\columnwidth]{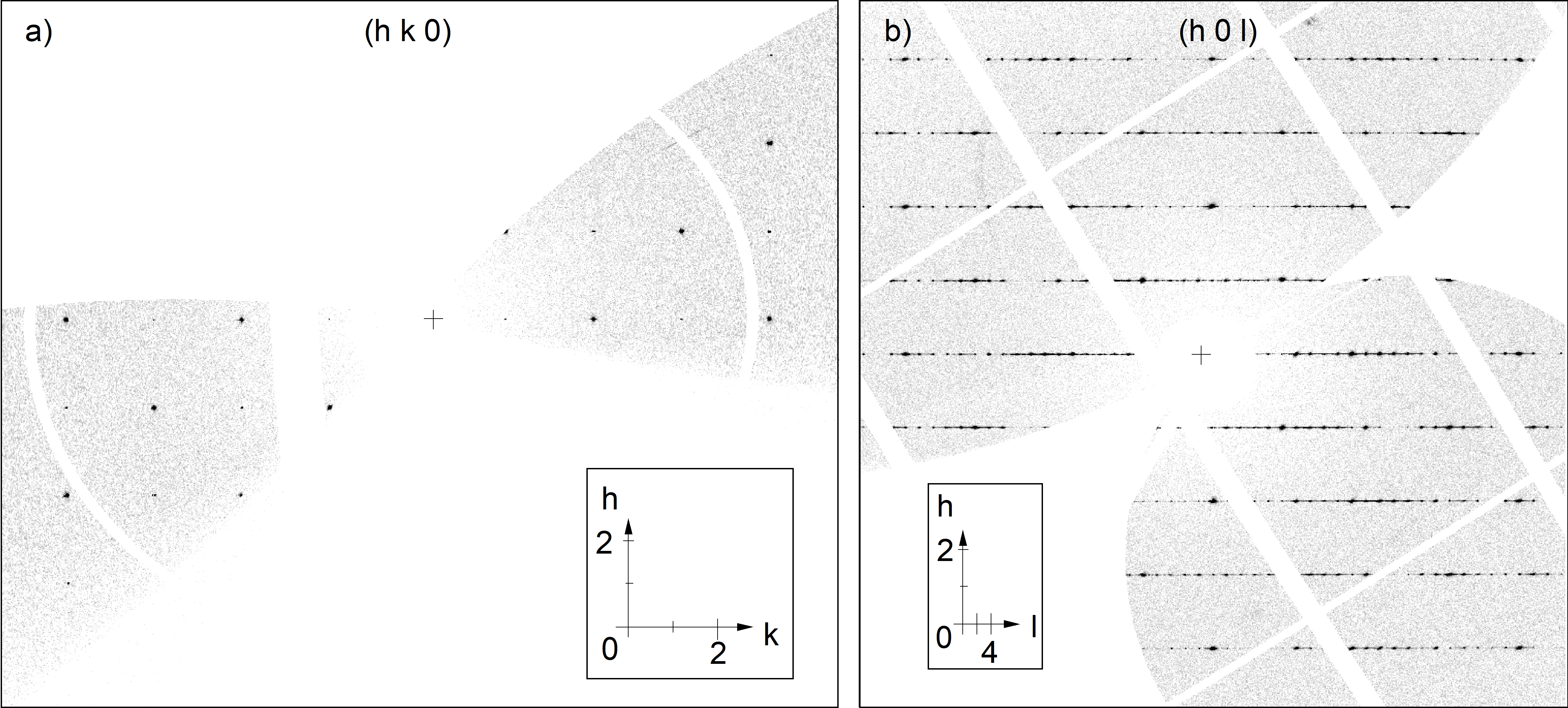}
\par\end{centering}
\caption{Reconstructed maps of XRD intensities of an annealed La$_3$Ni$_2$O$_{7}$ crystal for (a) the $(h, k, 0)$ plane and (b) the $(h, 0, l)$ plane. The cross symbol in the center of the maps marks the point (0, 0, 0). The insets in (a) and (b) indicate the spacing in reciprocal lattice units along the $h$ and $k$ and the $h$ and $l$ directions, respectively. The XRD data was acquired with synchrotron radiation ($\lambda = 0.22290$ {\AA}) at room temperature and an applied pressure of 0.7 GPa. The orthorhombic $Fmmm$ unit cell was used for the reconstruction.
}
\label{fig:XRDmaps}
\end{figure*}

\begin{figure*}[h]
 \begin{centering}
\includegraphics[width=1\columnwidth]{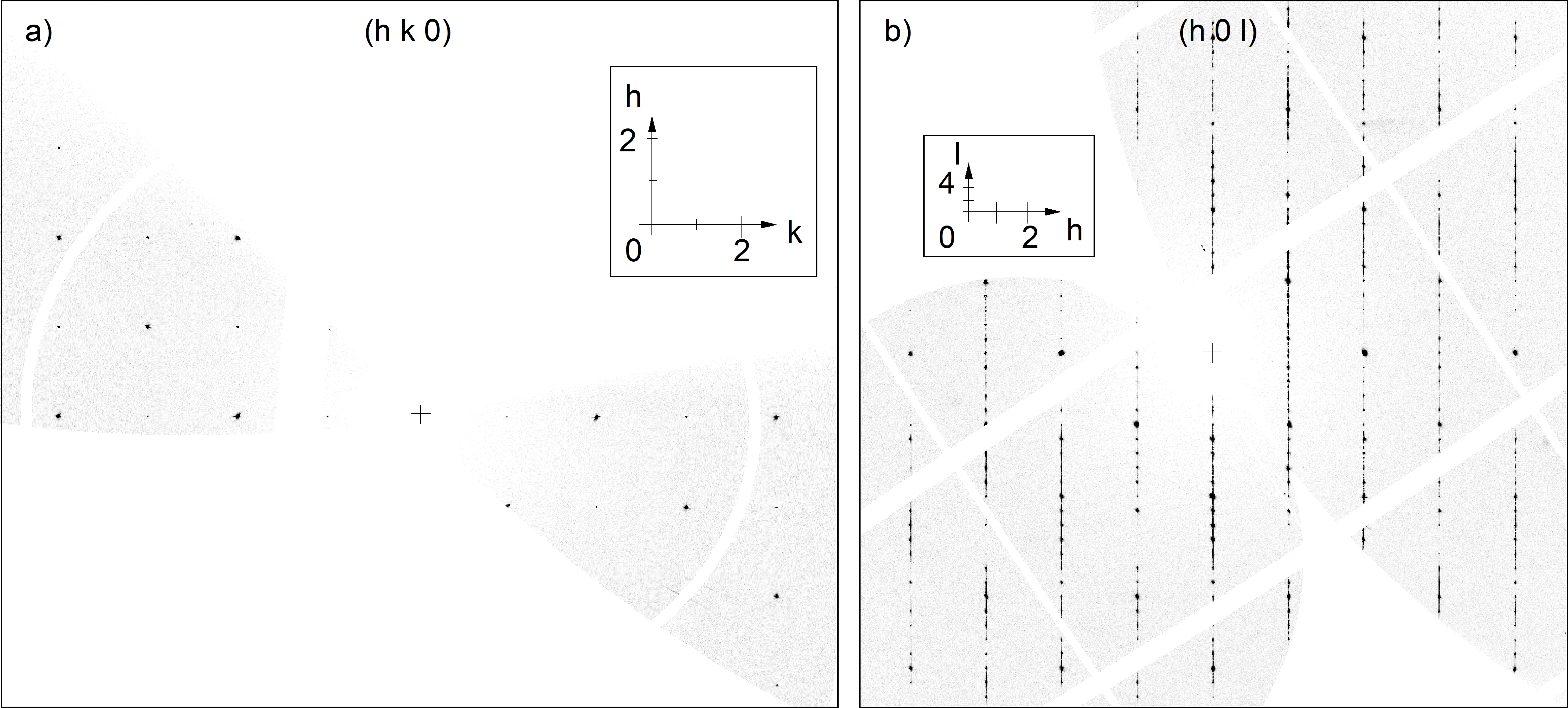}
\par\end{centering}
\caption{Reconstructed maps of XRD intensities of an annealed La$_3$Ni$_2$O$_{7}$ crystal for (a) the $(h, k, 0)$ plane and (b) the $(h, 0, l)$ plane. The data was acquired with synchrotron radiation ($\lambda = 0.22290$ {\AA}) at room temperature and an applied pressure of 16 GPa. The tetragonal $P4/mmm$ unit cell was used for the reconstruction.
}
\label{fig:XRDmaps2}
\end{figure*}

\begin{figure*}[h]
 \begin{centering}
\includegraphics[width=1\columnwidth]{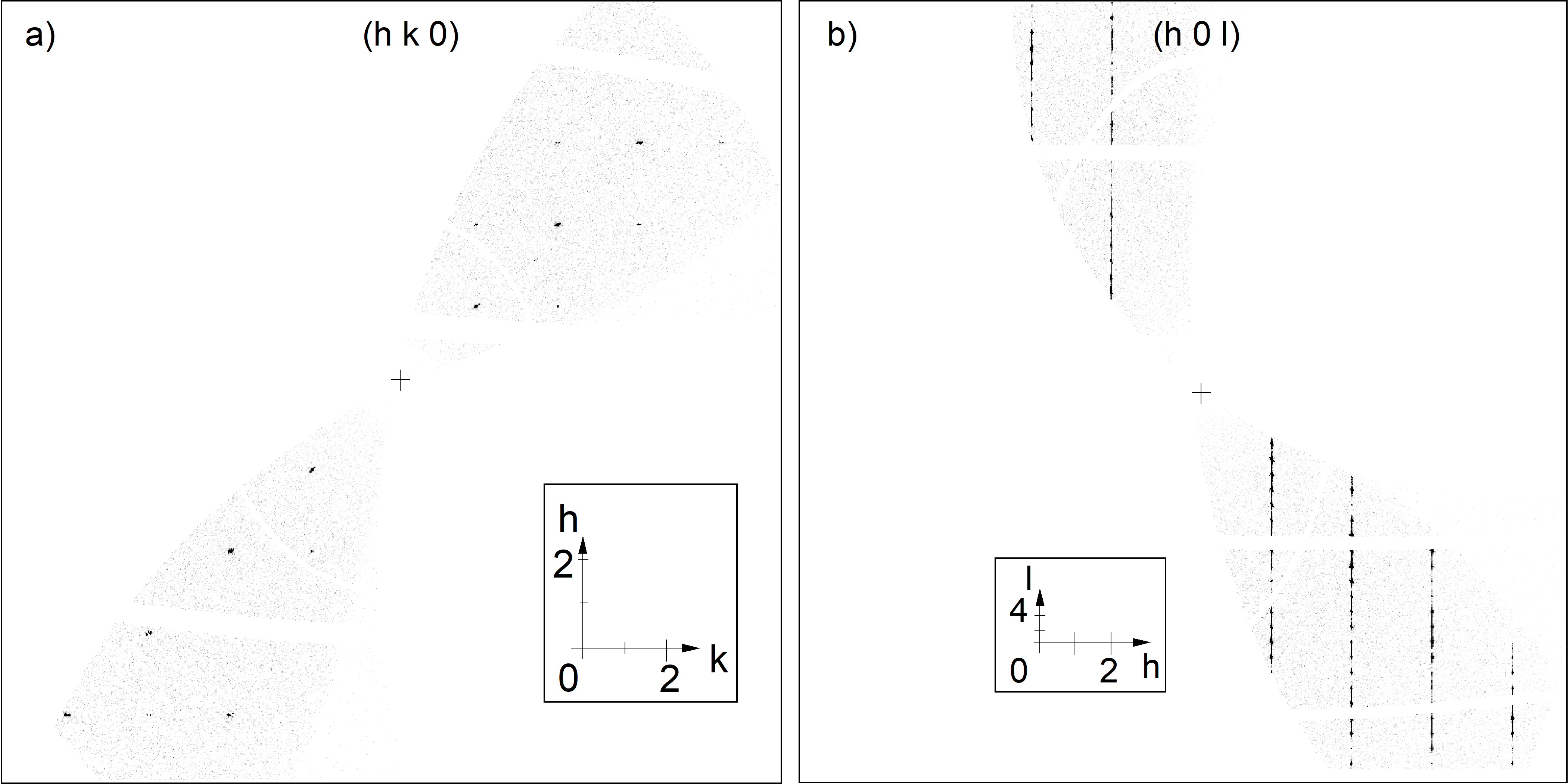}
\par\end{centering}
\caption{Reconstructed maps of XRD intensities of an as-grown La$_3$Ni$_2$O$_{6.83}$ crystal for (a) the $(h, k, 0)$ plane and (b) the $(h, 0, l)$ plane. The data was acquired with synchrotron radiation ($\lambda = 0.37851$ {\AA}) at room temperature and an applied pressure of 1.5 GPa. The tetragonal $P4/mmm$ unit cell was used for the reconstruction.
}
\label{fig:XRDmaps3}
\end{figure*}

\begin{figure*}[h]
 \begin{centering}
\includegraphics[width=1\columnwidth]{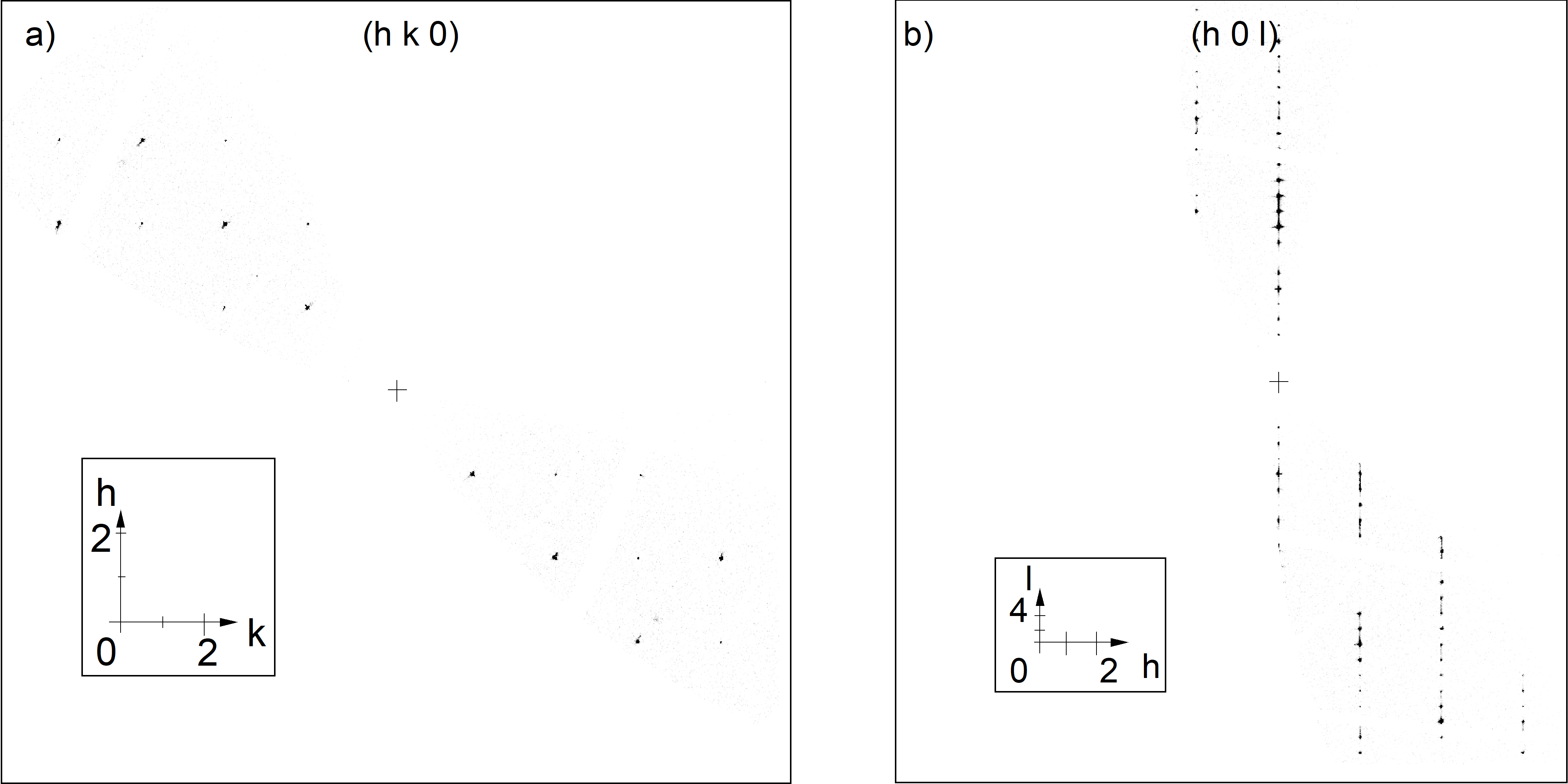}
\par\end{centering}
\caption{Reconstructed maps of XRD intensities of an as grown La$_3$Ni$_2$O$_{6.83}$ crystal for (a) the $(h, k, 0)$ plane and (b) the $(h, 0, l)$ plane. The data was acquired with synchrotron radiation ($\lambda = 0.37851$ {\AA}) at room temperature and an applied pressure of 14.9 GPa. The tetragonal $P4/mmm$ unit cell was used for the reconstruction.
}
\label{fig:XRDmaps4}
\end{figure*}

\begin{figure*}[h]
 \begin{centering}
\includegraphics[width=0.8\columnwidth]{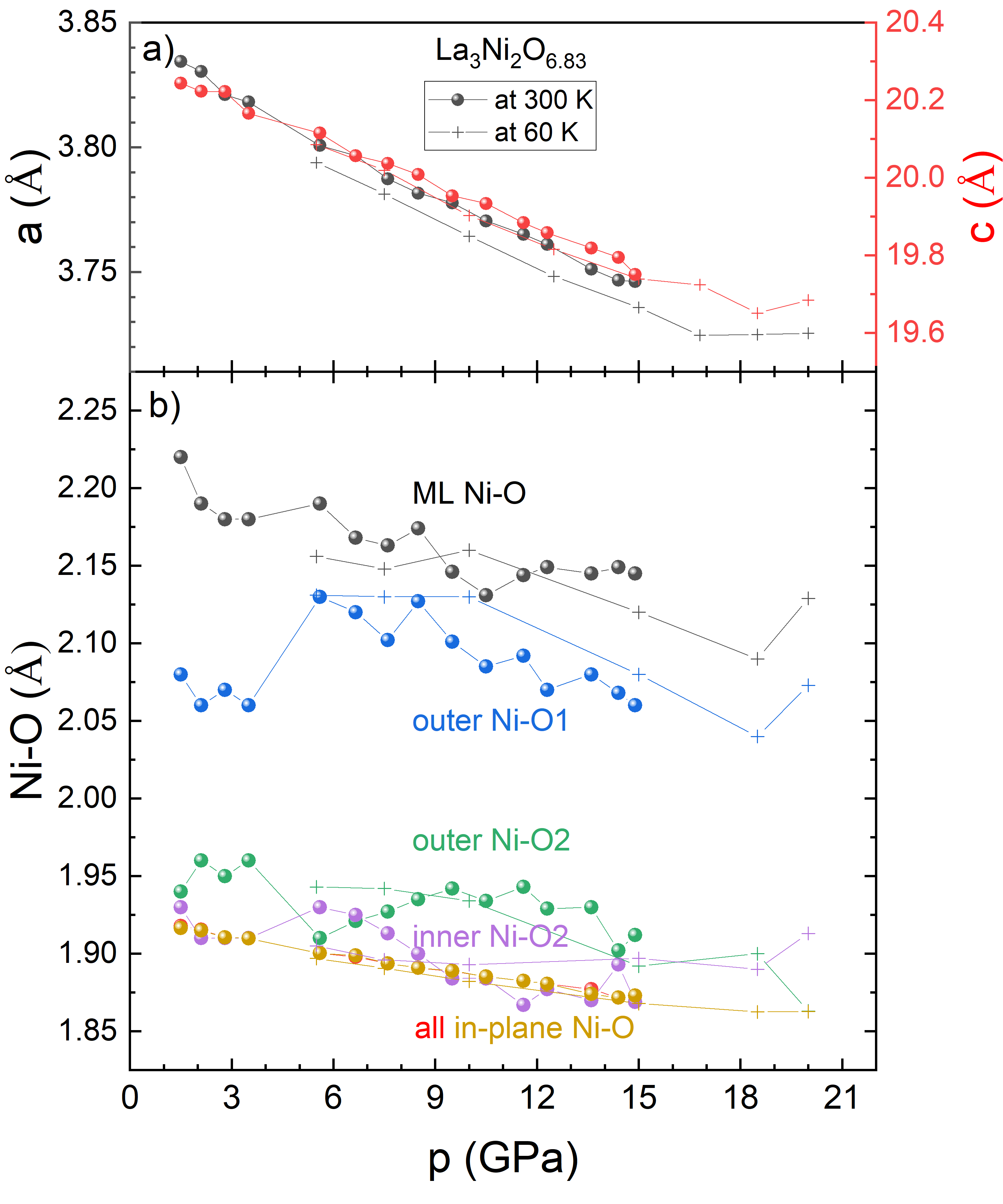}
\par\end{centering}
\caption{(a) Evolution of the lattice constant of an as grown La$_3$Ni$_2$O$_{6.83}$ with pressure both measured at 300 K (round symbol) and 60 K (cross symbol). (b) Evolution of the Ni-O bond distances of an as grown La$_3$Ni$_2$O$_{6.83}$ with pressure refined with $P4/mmm$ cell.
}
\label{XRDo6p83}
\end{figure*}

\begin{figure*}[tb]
 \begin{centering}
\includegraphics[width=0.8\columnwidth]{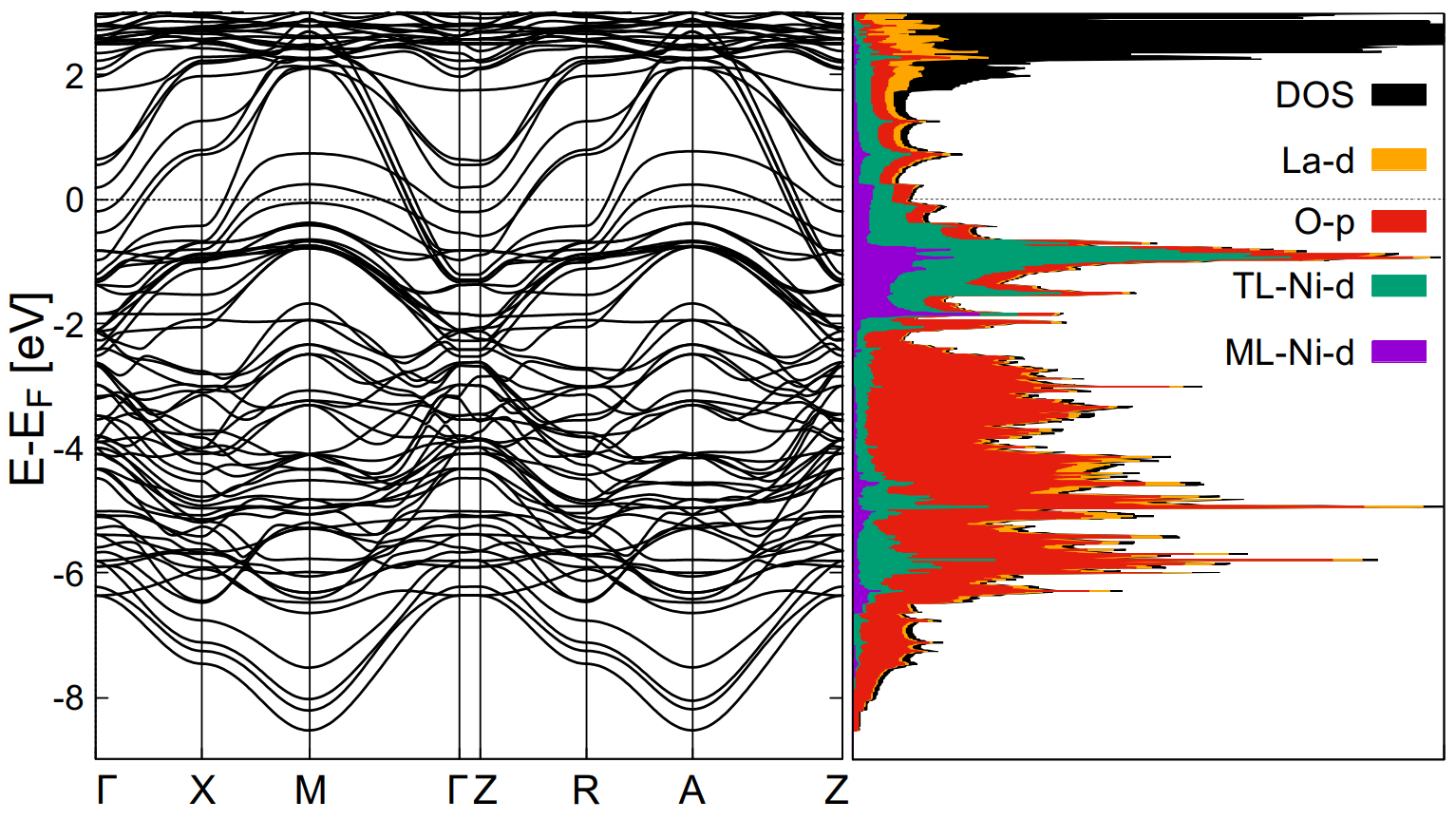}
\par\end{centering}
\caption{Left: Complementary plot of the DFT results for La$_3$Ni$_2$O$_7$ in the high-pressure $P/4mmm$ phase of the ML-TM structure according to the XRD refinement for 12.3 GPa pressure. The band structure is shown along a full path across the Brillouin zone, traversing also the $k_{z}=0.5$ plane, which was not shown in the Fig~4 of the main text. An extended energy range below the Fermi level is also shown, encompassing all O $p$ states. Right: The partial density of
states (DOS), plotted cumulatively.
}
\label{fig:DFT_large-bands}
\end{figure*}

\begin{figure*}[tb]
 \begin{centering}
\includegraphics[width=0.5\columnwidth]{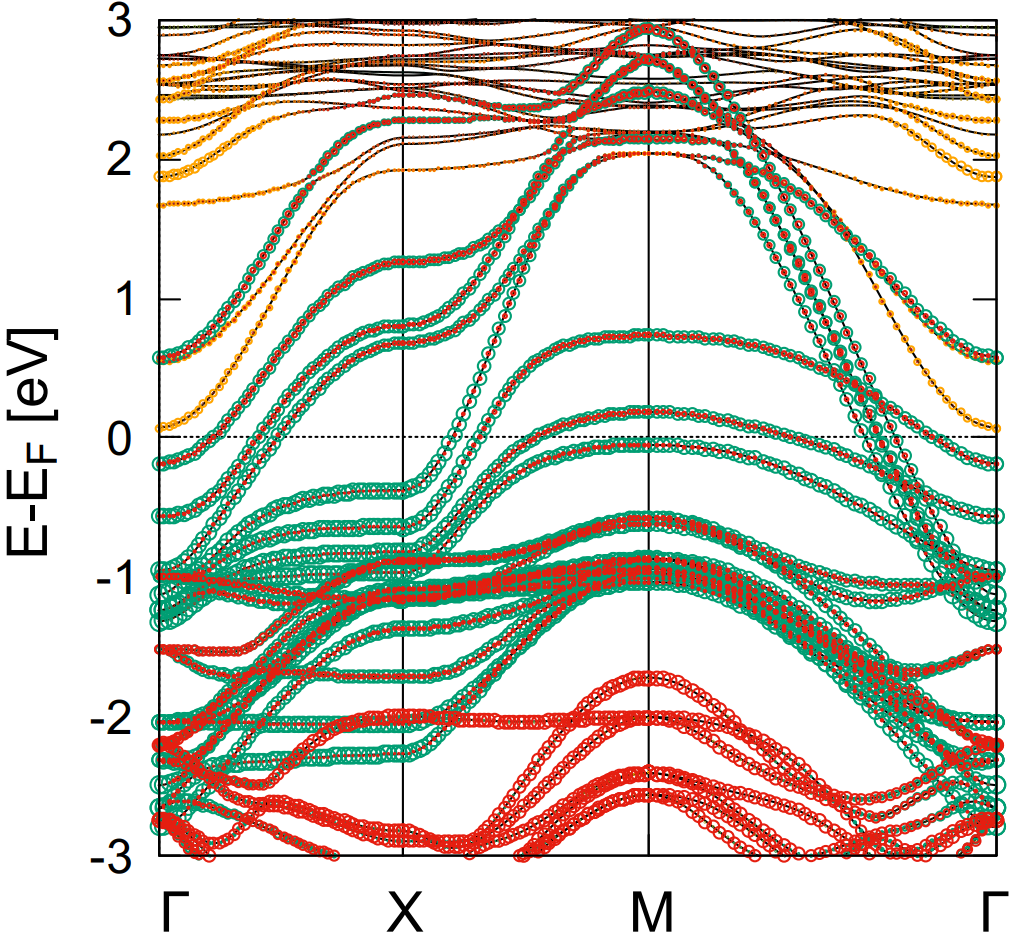}
\par\end{centering}
\caption{Complementary plot of the band structure in fatband representation, illustrating the contributions of La $d$ (orange), Ni $d$ (green), and O $p$ (red) states in a broad energy window around the Fermi level. Around the $\Gamma$ point, the first band right above the Fermi level exhibits La $d$ character without notable O $p$ and Ni $d$ contributions. The Ni $d$ bands mostly exhibit admixed O $p$ contributions. 
}
\label{fig:La_fatbands}
\end{figure*}

\begin{figure*}[tb]
 \begin{centering}
\includegraphics[width=1.0\columnwidth]{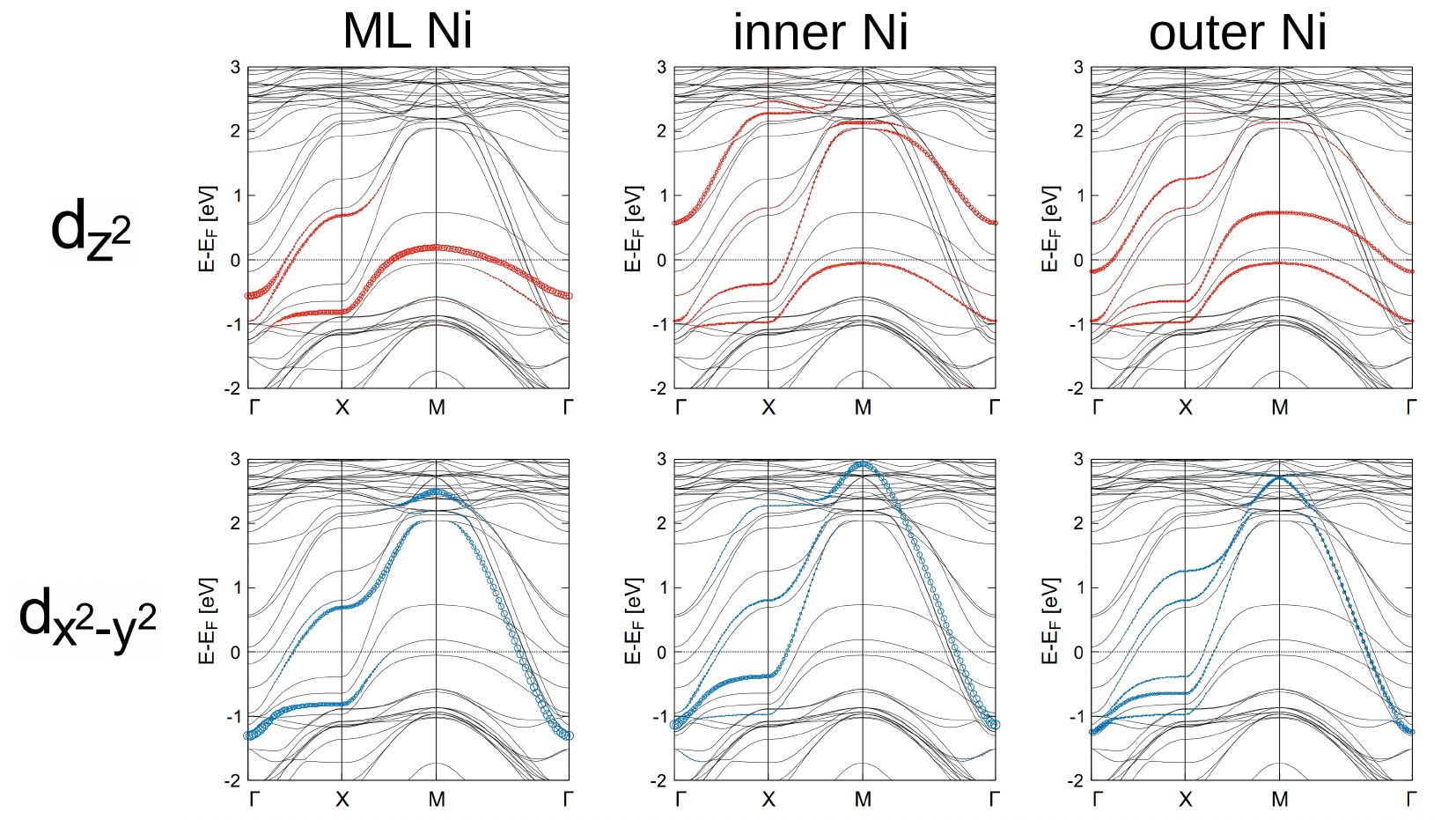}
\par\end{centering}
\caption{Complementary plots showing the Ni $e_g$ bands in fatband representation, disentangled according to the orbital states ($d_{z^2}$- and $d_{x^2-y^2}$) and Ni sites within the structural units (ML Ni, inner Ni, and outer Ni).
}
\label{fig:DFT_fatbands}
\end{figure*}

\end{widetext}
\end{document}